\documentclass[onecolumn]{mn2e}
\usepackage{graphicx}

\newcommand{\msat}{M_{\rm{sat}}}
\newcommand{\mhost}{M_{\rm{host}}}

\newcommand{\rvir}{r_{\rm{vir}}}

\newcommand{\tdf}{\tau_{\rm merge}}

\def\kms{\>{\rm km}\,{\rm s}^{-1}}

\def\Msun{\>{\rm M_{\odot}}}

\newcommand{\gtsim}{\mathrel{\hbox{\rlap{\lower.55ex \hbox {$\sim$}}
                   \kern-.3em \raise.4ex \hbox{$>$}}}}
\newcommand{\ltsim}{\mathrel{\hbox{\rlap{\lower.55ex \hbox {$\sim$}}
                   \kern-.3em \raise.4ex \hbox{$<$}}}}

\title{Black Hole Growth from Cosmological N-body Simulations}

\author[Miroslav Micic, Kelly Holley-Bockelmann, Steinn Sigurdsson]
{Miroslav Micic$^1$\thanks{E-mail: m.micic@vanderbilt.edu, k.holley@vanderbilt.edu,
steinn@astro.psu.edu}, Kelly Holley-Bockelmann$^1$, Steinn Sigurdsson$^2$ \\
$^1$ Department of Physics \& Astronomy, Vanderbilt University  \\
$^2$ Department of Astronomy \& Astrophysics, Pennsylvania State University  \\
}

\begin{document}
\maketitle

\begin{abstract}

We use high resolution cosmological N-body simulations to study the 
growth of intermediate to supermassive black holes (SMBH) from redshift 49 to
zero. Our cosmological volume and mass resolution is small enough that
we can track the growth of black holes from the seeds of population III
stars to black holes in the range of $10^3 < M < 10^7$ solar mass black holes
-- not quasars, but rather intermediate mass black holes to low-mass SMBHs. 
These lower mass black holes are the primary observable for the Laser 
Interferometer Space Antenna (LISA), and may remain the most common black hole
in the universe. The large-scale dynamics of the black holes are followed
accurately within the simulation down to scales of 1 kpc; thereafter, we
follow the merger analytically from the last dynamical friction phase
to black hole coalescence. We find that the merger rate of these black holes
is R$\sim$ 25 yr$^{-1}$ between 8 $\leq$ z $\leq$ 11 and R = 
10 yr$^{-1}$ at z=3. Before the merger occurs the incoming IMBH may
be observed with a next generation of X-ray telescopes as a ULX source 
with a rate of about $\sim$ 3 - 7 yr$^{-1}$ for 1 $\leq$ z $\leq$ 5.
When we include the lowest predicted efficiency of Pop III star formation, the
black hole merger rate decreases to R $\sim$ 2 - 3 yr$^{-1}$ which is still
a reliable LISA source. 

We develop an analytic prescription that captures the most important 
black hole growth mechanisms: galaxy merger-driven gas accretion and 
black hole coalescence. This prescription is based on sophisticated 
gas-rich galaxy merger simulations that include feedback and star formation.
Using this simple prescription, we find that we can grow an analogue of  the 
Milky Way black hole and dark matter halo by redshift zero. 
In our volume, this supermassive black hole was in place with most of 
its mass by z = 4.7, and most of the growth was driven
by gas accretion excited by major mergers.

Hundreds of black holes have failed to coalesce with the SMBH
by $z=0$, some with masses of $10^4 M_\odot$. As these black holes
orbit within the dark matter halo, they may have luminosities up to 
$\sim 30000 L_\odot$ if the halo is suffused with tenuous hot gas. These 
bright X-ray sources can easily be observed with Chandra at distances of 
$\sim$ 100 kpc.

\end{abstract}


\section{INTRODUCTION}

Supermassive black holes, with masses of 
$10^6 M_\odot \leq M \leq 10^{10} M_\odot$ are widely believed to 
dwell at the centers of elliptical galaxies and spiral bulges 
(e.g. Kormendy $\&$ Richstone 1995), and the most massive of these 
black holes are tied to quasar phenomena at high redshift 
(Greenstein $\&$ Matthews 1963). There is abundant evidence 
that when a SMBH is in place, it transforms the structure and evolution 
of the galaxy, from powering active galactic nuclei at high 
redshifts (Rees 1984, Alexander et al. 2005, Fan 2005), to regulating 
star formation throughout the galaxy (Di Matteo et al. 2005, 
Cox et al. 2008), to scouring the galactic nucleus of stars 
during SMBH mergers (Ebisuzaki et al. 1991, Quinlan 1996, Makino 1997, 
Milosavljevic $\&$ Merritt 2001, Volonteri et al. 2003).

This deep connection between the evolution of SMBHs and galaxies
is perhaps best encapsulated in a remarkable correlation
between the SMBH mass and the velocity dispersion of the host 
spheroid (Gebhardt et al. 2000, Ferrarese $\&$ Merritt 2000, 
Tremaine et al. 2002). This $M_{\rm BH} - \sigma$ correlation 
appears to be intrinsically perfect, at least for a sample of nearby 
bright spiral and elliptical galaxies with clear dynamical SMBH 
signatures. There is even a suggestion that the $M_{\rm BH} - \sigma$ 
relation may hold for black holes on a much smaller mass 
scale of $10^3 < M_{\rm BH} < 10^5 M_\odot$ -- the intermediate mass 
black holes, or IMBHs (Filippenko $\&$ Ho 2003, Barth et al. 2004, 
Gebhardt et al. 2005, Green $\&$ Ho 2004).

In searching for the root cause of this galaxy-SMBH connection, a handful of  
SMBH mass-'host property' correlations have emerged. Many of these 
correlations, admittedly, can simply be byproducts of the 
$M_{\rm BH}-\sigma$ relation, since the velocity dispersion of a 
galaxy is a measure of its underlying potential. 
However, several observational and theoretical studies have linked the 
SMBH mass with the mass of the host dark matter 
halo (Ferrarese 2002, Baes et al. 2003, Shankar et al. 2006). 
This relation is a boon to theorists because many of the leading 
explanations of SMBH birth and growth are driven by hierarchical structure 
formation (Hopkins et al. 2005, Wyithe $\&$ Loeb 2005, Granato et al. 2001,
Menou et al. 2001, Adams et al. 2001, Monaco et al. 2000, Silk $\&$ Rees 1998,
Haehnelt et al. 1998, Haehnelt $\&$ Kauffmann 2000, Cattaneo et al. 1999, 
Loeb $\&$ Rasio 1994), and are therefore tied to the mass of the dark matter halo.

In the current picture of SMBH assembly, the 
black hole begins life as a low mass ``seed'' black hole at high redshift. 
It's not clear, though,  when exactly these BH seeds emerge or
what mass they have at birth. SMBH seeds may 
have been spawned from the accretion of low angular momentum gas in 
a dark matter halo (Koushiappas et al. 2004, Bromm $\&$ Loeb 2003), 
the coalescence of many seed black holes within a 
halo (Begelman $\&$ Rees 1978, Islam et al. 2004), or from an IMBH formed, 
perhaps, by runaway stellar collisions (Portegies Zwart et al. 2004, 
Miller $\&$ Colbert 2004, van der Marel 2004). However, the most likely
candidates for SMBH seeds are the remnants that form from the first 
generation of stars sitting deep within dark matter 
halos (Madau $\&$ Rees 2001, Heger et al. 2003, Volonteri et al. 2003, 
Islam et al. 2003, Wise $\&$ Abel 2005) -- so called Population III stars. 
With masses $< 10^3 \Msun$, these relic seeds are predicted to lie near 
the centers of dark matter halos between $z\sim 12-20$ (Bromm et al. 1999, 
Abel et al. 2000, 2002). Structure 
formation dictates that dark matter halos form in the early universe
and hierarchically merge into larger bound objects, so naturally as 
dark matter halos merge, seed black holes sink to the center through dynamical
friction and eventually coalesce. Dark matter halo mergers become 
synonymous, then, with black hole mergers at these masses and redshifts.
This means that although the seed formation stops at z$\sim$12 as 
Population III supernovae rates drop to zero (Wise $\&$ Abel 2005), 
SMBH growth continues as dark matter halo mergers proceed to low redshifts.

Gas accretion is thought to play a critical role in 
fueling the early stages of black hole growth (David et al. 1987, 
Kauffmann $\&$ Haehnelt 2000, Merloni 2004), and this may explain 
the tightness of the $M_{\rm BH}-\sigma$ relation (Burkert $\&$ Silk 2001,
Haehnelt $\&$ Kauffmann 2000, Di Matteo et al. 2005, Kazantzidis et al. 2005, 
Robertson et al. 2006). Since high redshift galaxies are thought to be 
especially gas-rich, each merger brings a fresh supply of gas to the center 
of the galaxy, and new fuel to the growing supermassive black hole 
(Mihos $\&$ Hernquist 1994, Di Matteo et al. 2003). From a combination 
of gas accretion and binary black hole coalescence, it is thought that 
these Pop III-generated seeds may form the SMBHs we observe 
today (Soltan 1982, Schneider et al. 2002).

During a galaxy merger, each black hole sinks to the center of the new galaxy 
potential due to dynamical friction and eventually becomes bound as 
a binary (Kazantzidis et al. 2005; Escala et al. 2005). Dynamical 
friction then continues to shrink the orbit until the binary is hard 
(i.e, the separation between each black hole, $a_{\rm BBH}$, is such that the 
system tends to lose energy during stellar encounters) (Heggie et al. 2007).
Thereafter, further decay is
mediated by 3-body scattering with the ambient stellar background
until the binary becomes so close that the orbit can lose energy
via gravitational radiation. In studies of static, spherical
potentials, it may be difficult for stellar encounters alone to 
cause the binary to transition between the 3-body scattering phase
and the gravitational radiation regime (Milosavljevic $\&$ Merritt 2003).
 However, in gas-rich or
non-spherical systems, the binary rapidly hardens and coalesces into one
black hole, emitting copious gravitational radiation in the process 
(Mayer et al. 2007, Kazantzidis et al. 2005, Berczik et al. 2006; 
Sigurdsson 2003; Holley-Bockelmann $\&$ Sigurdsson 2006).

In our previous work, we calculated the cosmological merger rate for 
black holes between 200 - $3 \times 10^7 \Msun$ from redshift 20-0 
(Micic et al. 2007). Our approach
combined high-resolution, small-volume cosmological N-body simulations
with analytic prescriptions for the dynamics of merging black holes below our
resolution limit; this allowed us to explore different black hole growth 
mechanisms and seed formation scenarios while also accurately simulating the
rich and varied merger history of the host dark matter halos. However, our
original work treated the black hole merger timescales rather optimistically;
the black holes were said to merge immediately after the dark matter halos 
merge. In addition, we assumed that black hole seed formation is prolific, 
occurring in every dark matter halo capable of hosting a Population III star. 
The merger rates from Micic et al. 2007, then,
may be considered upper limits for a given black hole growth scenario.

It is expected that only a fraction of first 
stars collapse into massive black holes. Calculating this fraction would be
trivial if the POP III initial mass function is known. Unfortunately,
the initial mass function of the first stars is still very uncertain, 
mostly because the primordial fragmentation process is still very poorly 
understood (Nakamura $\&$ Umemura 2001; Omukai $\&$ Yoshii 2003). If the 
initial mass function for the first stars were top heavy 
(Schneider et al. 2002), $\sim$ 6 $\%$
of the total mass in first stars would go into black holes after supernovae
collapse. Since half of the mass of each individual first star is 
ejected during a supernovae (Heger $\&$ Woosley 2002), $\sim$ 12 $\%$ of 
first stars produce a black hole. Hence, the fraction of the first stars 
that produce massive black holes could be as low as 10 $\%$, 
f$_{\rm POPIII}$ = 0.1 (Schneider et al. 2002). This will have a direct 
consequence on the occupation fraction of halos with seed black holes at 
high redshift.

Unless the black hole binaries stall at the final parsec, the longest 
timescale governing the coalescence of two black holes occurs when  
the host galaxies themselves are still merging. 
Here, the dynamical evolution of two merging galaxies is driven by
the combined effect of dynamical friction that brings the
less massive galaxy (satellite) to the center of the larger halo (primary),
and the tidal interaction that strips mass from the satellite
(e.g. Richstone 1976, Aguilar $\&$ White 1986, 
Holley-Bockelmann $\&$ Richstone 1999, Taffoni et al. 2003) and 
further delays the merger. If the dynamical friction timescale
is longer than a Hubble time, the black holes carried by their host
galaxies will not be brought close enough to form a binary. This 
delay in the merger timescale can certainly reduce black hole merger rates.

In this paper, we continue to use this hybrid method to study how our 
predictions of black hole assembly changes when the effects of black hole 
seed formation efficiency and dynamical friction are introduced, and what 
effect these processes have on merger rates. We track the assembly of black 
holes over a large range of final masses -- from seed black holes to SMBHs -- 
over widely varying dynamical histories. We also expand on a prescription 
to model the growth of the black hole from a seed to a supermassive black 
hole as a function of the merger history; this prescription is physically 
motivated by simulations of star formation and feedback, captures the 
two most important mechanisms by which supermassive black holes assemble, 
and is very easy to implement in cosmological dark matter-only simulations. 
With this improved treatment of black hole
dynamics and accretion, we derive merger rates of dark matter halos and
black holes as a function of mass ratio and redshift. These will
be important observables for the Laser Interferometric Space Antenna (LISA),
a planned space mission to detect gravitational waves, set to launch in 
the next decade.

We describe our method in section 2 and introduce the black hole growth 
prescription. In section 3, we present the results for both the growth of 
SMBH in our volume and for the properties of rogue black holes; this 
section includes black hole merger rates, dark matter halo merger rates 
in this low mass regime, and our derived black hole-host halo relation
on this low mass end. We discuss the implications of our results and 
future work in section 4.

\section{METHOD}

In this section, we give an overview for how we simulate our 
small cosmological volume, construct the dark matter halo merger trees, 
and model the black hole physics.

\subsection{Simulation Setup and Dark Matter Halo Merger Tree}

Using GADGET (Springel et al. 2001), we performed a high-resolution 
cosmological N-body simulation within a comoving 10 Mpc$^3 $ section of 
a $\Lambda$CDM universe ($\Omega_{\rm M}$=0.3, $\Omega_{\Lambda}$=0.7, 
$\sigma_8$=0.9 and h=0.7) from $z=40$ to $ z=0$.  We are using 
WMAP1 (Bennett et al. 2003) cosmological parameters in this study to
compare with our previous work; however, at the time of this paper's 
submission, simulations of several small volumes are underway with 
WMAP3 parameters (Spergel et al. 2007). A detailed description of this 
simulation can be found in Micic et al. 2006. Briefly, however, we identify 
high density region within a low-resolution cosmological volume that 
at $z=0$ hosts a dark matter halo comparable in size to the halo hosting 
the Milky Way Galaxy. We then refine a sphere of 2 Mpc around the halo of 
interest to simulate at a higher resolution with 4.9$\times10^6$ 
high-resolution particles (softening length 2 kpc comoving). The rest 
of the box has 2.0$\times10^6$ low-resolution particles (softening length 4 kpc
comoving). The mass of each high resolution particle in this simulation is
8.85$\times10^5$M$_\odot$, and the mass of each low-resolution particle is
5.66$\times10^7$M$_\odot$. In the post-simulation analysis, we identify 
dark matter structures, seed them with Population III black holes, and 
follow their merger history from redshift 20 to 0 by constructing numerical 
merger trees. The method we use is described in detail in Micic et al. 2007. 

In our hybrid method, we combine dark matter halo merger trees obtained in 
numerical simulations with an analytical treatment of the physical processes 
that arise in the dynamics of galaxy and black hole mergers. 
Our N-body approach stops with the creation of the halo 
merger tree. To define the structure of each dark matter halo within 
the n-body generated merger tree, we assume a 
Navarro, Frenk, $\&$ White (hereafter, NFW) density 
profile (Navarro, Frenk $\&$ White 1995). We set the parameters of a 
given NFW halo using the approach presented in Bullock et al. 2001, 
assuming the typical virial mass of a dark matter halo to be 
M$_{\rm typ}$ = 1.5 $\times$ 10$^{13}$ $\Msun$ at redshift 0 and 
that this mass varies with redshift as:
M$_{\rm typ}$(z) = M$_{\rm typ}$(z=0) / (1+z). The concentration, then, 
is defined as c$_{\rm vir}$ = 9 $\mu^{\alpha}$, where $\mu$ is the 
halo mass in units of the typical halo mass at the same redshift, 
$\mu$ = M$_{\rm halo}$/M$_{\rm typ}$, $\alpha$ = -0.13 for $\mu$ $\geq$ 0.2 
and $\alpha$ = -0.3 for $\mu$ $\leq$ 0.2. Concentration parameters for all 
dark matter halos in our merger tree as a function of halo mass are 
presented in figure 1. For each merger, we tag the more massive halo as 
the primary, with mass M$_{\rm p}$, and the less massive halo as the 
secondary or satellite halo with mass M$_{\rm s}$. 
Note that the properties of dark matter halos in 
the mass range of our simulation (
10$^{6.5}\Msun$ $\leq$ M $\leq$ 10$^{12}$ $\Msun$) have not 
been studied at high redshift (z $\geq$ 3).

\subsection{Black Hole Seed Formation}

In the current picture of the evolution of very low metallicity stars,
those with an initial mass 
30 $\Msun$ $\leq$ M$_{\rm POP III}$ $\leq$ 140 $\Msun$ and
M$_{\rm POP III}$ $\geq$ 260 $\Msun$ collapse directly into a black hole
(Heger $\&$ Woosley 2002). Due to copious mass loss from winds, 
stars $30 \leq M_{\rm POP III}$ form a neutron
star or white dwarf, and stars with 
140 $\Msun$ $\leq$ M$_{\rm POP III}$ $\leq$ 260 $\Msun$
go through pair instability supernovae leaving no remnant behind.
The initial mass function (IMF) of these first stars is unclear, though, in 
part due to a lack of understanding of how
primordial molecular hydrogen gas clouds fragment. It is expected 
that the IMF may be top heavy. This is
because the accretion rate onto the first stars is large
(Omukai $\&$ Nishi 1998, Ripamonti et al. 2002)
and since dust grains are absent in primordial H$_2$ gas clouds, the
radiative feedback from the forming star is not strong enough
to halt further gas accretion (Omukai $\&$ Palla 2003, Bromm $\&$ Loeb 2004).
The latest constraint of the first star's maximum mass comes from
modeling the structure of the accretion flow and the evolution of
protostars, and sets the maximum mass to $\sim$ 300 $\Msun$. It is unlikely
that the Pop III IMF is anything like a Salpeter IMF (power law with
-2.35 slope), a canonical IMF used to model metal-enriched stellar populations.

The fraction of the first stars that produce massive black holes
could be as low as 10 $\%$, f$_{\rm POPIII}$ = 0.1
(Schneider et al. 2002). We do note that 10$\%$ is a lower constraint and 
that much higher efficiency would not be unexpected. Since the formation of the
first stars is only limited, theoretically, by the minimum mass of the host
dark matter halo (Wise $\&$ Abel 2005), there is no expected
correlation between the masses of first stars and the masses of their 
host halos. We choose our seed black hole mass to be $ 200 \Msun$ in 
this work. In 100 realizations, we randomly select 10$\%$ of dark 
matter halos that can host seed black holes and remove the rest of 
the seeds from our merger tree. The effect of the initial occupation 
fraction on the black hole merger rates is direct and dramatic, as 
can be seen in figure 3.

\subsection{Dynamical Friction}

There are two important mechanisms that lead to the formation of a
massive black hole binary at the center of a galaxy. First, 
dynamical friction expedites the merger of two dark matter halos and 
later the merger of the galaxies they host. In this manner, 
merging galaxies can efficiently shepherd massive black holes to 
the center of the new system, roughly to the inner kiloparsec 
(see Colpi et al. 2007 for a review). Second, dynamical
 friction from the gas in the disk carries black holes deeper toward
the galactic center where they form binary and eventually merge (e.g. 
Begelman, Blandford $\&$ Rees 1980, Escala et al. 2005, 
Kazantzidis et al. 2005, Dotti et al. 2007). We model both effects as follows:

The time for a satellite to sink to the center of a primary can 
be approximated using Chandrasekhar dynamical friction (Binney $\&$ Tremaine 1987):

\begin{equation}
t_{\rm chandra}=\frac{1.17}{{\ln}{\Lambda}}\frac{{\rm r_{\rm circ}^{2} 
v_{c} {\epsilon^{\alpha}}}}{{\rm GM_{\rm s}}},
\end{equation}

\noindent where $\ln\Lambda$ is the Coulomb logarithm, $\ln\Lambda$ $\approx$ 
$\ln {\rm (1 + M_{\rm p}/M_{\rm s})}$. To define the satellite orbit, 
we adopt values suggested by numerical investigations 
(Colpi et al. 1999) and used in previous semi-analytical work (Volonteri et al. 2003), 
the eccentricity $\epsilon^{\alpha}$ = 0.8, and the circular velocity v$_{\rm c}$ 
is determined at r$_{\rm circ}$ = 0.6 r$_{\rm vir}$. 

Assuming that each merging galaxy carries a massive black hole at the 
center, t$_{\rm fric}$, is the merging time for massive black holes when 
all other processes (3-body scattering, gas dynamical friction, 
gravitational radiation, etc.) involved in the formation and later shrinking 
of the black hole binary are efficient and fast. Due to tidal stripping 
and possible resonant interactions, simulations have shown that the 
Chandrasekhar formula underestimates the merger time, especially in 
the case of minor mergers (e.g. Holley-Bockelmann $\&$ Richstone 1999, 
Weinberg 1989). If this is true, then semi-analytic studies of black 
hole merger rates using a Chandrasekhar formalism for the merger 
time will overestimate the true number of black hole mergers.

In an effort to better parametrize dynamical friction, 
Boylan-Kolchin et al. 2008 used N-body simulations to study dark matter 
halo merging timescales, and confirmed that the Chandrasekhar formalism 
does underestimate the merger time, by a factor of 
$\approx$ 1.7 for M$_{\rm p}$/M$_{\rm s}$ $\approx$ 10 and a
factor of $\sim$ 3.3 for M$_{\rm p}$/M$_{\rm s}$ $\approx$ 100. 
They propose a fitting formula that accurately predicts the timescale 
for a satellite to sink from the virial radius to the host halo center: 

 \begin{equation}
   \frac{\tdf}{\tau_{\rm dyn}} = A \, {(\mhost/\msat)^b \over \ln(1+\mhost/\msat)}
  \exp\left[c \, {j \over j_c(E)} \right] \, \left[{r_c(E) \over \rvir} \right]^d,
\end{equation}
where b = 1.3, c = 1.9, d = 1, A = 0.216, circularity 
j / j$_c$(E) = 0.5, r$_c$(E) / r$_{\rm vir}$ = 0.65, as defined in Boylan-Kolchin et al. 2008. 
The dynamical time, t$_{\rm dyn}$, is given at virial radius as:

\begin{equation}
   \tau_{\rm dyn}\equiv \frac{\rvir}{V_c(\rvir)}
           =\left( \frac{\rvir^3}{G\mhost} \right)^{1/2} \,,
\end{equation}
where V$_{\rm c}$(r$_{\rm vir}$) = (GM$_{\rm vir}$/R)$^{1/2}$.

Figure 2 shows the dark matter halo merger timescale for each pair of 
merging halos for each of these two dynamical friction estimates. We 
have calculated the dark matter halo merger rate with both 
Chandrasekhar dynamical friction formula and the Boylan-Kolchin 
numerical fit. After showing the variance in dark matter halo merger 
rates from using the two expressions for the dynamical friction 
timescale (figure 3, discussed in section 3), we thereafter use
the Boylan-Kolchin timescale for all other calculations.

It is expected that in gas-rich galaxies, dynamical friction from 
the gas would bring two black holes close enough to form a binary 
whose orbit would shrink efficiently, passing quickly from a binary 
in the 3-body scattering phase to one emitting significant 
gravitational radiation (Escala et al. 2005, Kazantzidis et al. 2005, 
Dotti et al. 2007). Numerical simulations indicate that two black 
holes will sink from $\sim$ 1 kpc to form a binary with a separation 
of less than a parsec in $\sim$ 10 Myrs. 
We incorporate this physics by calculating the dynamical 
friction timescale from the virial radius to the 
inner kpc, and then assume that the two black holes merge 10 Myr 
afterward. In practice, the dynamical
friction timescale from equations 1 and 2 from the inner kpc 
to the bound binary stage is often of order
10 Myr; the power of this gas-rich assumption lies in that it 
entirely circumvents the so-called 'final-parsec' problem thought 
to exist for low mass ratio mergers of 
$10^{6.5} \Msun$ $\leq$ M$_{\rm BH}$ $\leq 10^8 \Msun$ 
within static, spherical, gas-poor galaxy models 
(e.g. Milosavljevic $\&$ Merritt 2003). We explicitly
assume that the black holes in our simulation do not stall at 
the final parsec before merger, and instead
are ushered efficiently into the gravitational radiation stage, 
where they coalesce; this assumption is 
verified even in the case of gas-poor galaxy models as long as 
the model is not spherical or in equilibrium 
(e.g. Holley-Bockelmann $\&$ Sigurdsson 2006, Berczik et al. 2006, 
R. Spurzem, private communication). Note that this implies that if 
we assume that each halo initially carries a black hole at its center, 
and that each host galaxy is gas-rich,
figure 2 also estimates the time for a black hole to sink from the
virial radius to the center of the host galaxy, become a bound 
black hole binary, and inspiral due to gravitational radiation.

In our initial work (Micic et al. 2007), mergers of dark matter halos 
trigger the immediate merger of the black holes they are hosting. 
In this paper, subsequent mergers of the central black holes are 
delayed to account for dynamical friction of the halos and the 
black holes within the galaxy. Black holes will not merge if 
their merger time is larger than a Hubble time, and in that case, 
we advance the black hole position within the primary halo 
at each timestep. Knowing the dynamical friction timescale for 
each merger, we postpone the black hole mergers by moving them down the merger
tree. For the final kpc, we assume that the ambient gas and non-sphericity 
will cause two black holes to coalesce within 10 Myrs.

\subsection{SMBH Growth Prescription}

The SMBH in our model grows through a combination of black hole 
mergers and gas accretion. To better separate the effects
of gas accretion on the black hole, we include a dry growth scenario, 
where the black hole grows through mergers only. To review the approach 
we have taken in a previous paper, we assumed that major galaxy 
mergers would funnel gas to the black hole in each of the progenitors 
and activate an Eddington-limited growth phase for a Salpeter time. Here, 
the black hole mass would grow as: 
M$_{\rm BH}$(t) = M$_{\rm BH,0}$(t$_0$) exp($\Delta$t/t$_{\rm sal}$), 
where $\Delta$t = t - t$_0$, $t_{\rm sal} \equiv  \epsilon M_{\rm BH} c^2 / [(1-\epsilon) L]$, $\epsilon$ 
is the radiative efficiency, L is 
the luminosity, and c is the speed of light; in this picture the black hole
mass would roughly double in 40 Myr (Hu et al. 2006) . We distinguished 
two cases depending on the mass ratio of merging dark matter halos. The 
first is a more conservative criterion that allows black holes to accrete 
gas if the mass ratio of the host dark matter halo is less than 4:1. 
The second case sets an upper constraint on the final black 
hole mass by allowing gas accretion as long as the merging dark matter 
halos have a mass ratio less than 10:1 -- on the cusp 
of what is considered a minor merger. Since our black holes merged 
promptly after the halos merged,
the accretion timescale and efficiency for major mergers was the same 
regardless of the mass ratio or redshift.

In this paper, we continue to model the black hole growth as one of 
extended gas accretion excited by major mergers. At high redshift, 
this is likely a good assumption, though note that at low redshift when
mergers are infrequent, secular evolution, such as bar instabilities, 
may dominate the gas (and therefore black hole) accretion. Integrated over 
the whole of a black hole lifetime, though, this
major merger-driven accretion is likely to be the dominant source of 
gas inflow. Since the black hole growth is so strongly dependent on what 
fuel is driven to the center during galaxy mergers,
it is important to characterize this merger-driven gas inflow, including 
the critical gas physics that may inhibit or strengthen this nuclear supply.
We are motivated by a recent suite of numerical simulations that include 
radiative gas cooling, star formation, and stellar feedback to study the 
starburst efficiency for unequal mass ratio galaxy mergers (Cox et al. 2008), 
which finds that the gas inflow depends strongly on the mass ratio of the 
galaxy (see also, e.g.,  Hernquist 1989, Mihos $\&$ Hernquist 1995).
This study parametrizes the efficiency of nuclear star formation 
(i.e. gas supply and inflow), $\alpha$, as a function of galaxy mass ratio:

\begin{equation}
{\alpha} = \Bigg({M_{\rm s} \over { M_{\rm p}}} - \alpha_0 \Bigg),
\end{equation}

\noindent where $\alpha_0$ defines the mass ratio below which there is no
 enhancement of nuclear star formation (i.e. gas inflow).  Here, the 
gas accretion efficiency has a maximum of 0.56 for 1:1 halo mergers and 
falls to zero at $\alpha_0$. This parametrization is insensitive to the 
stellar feedback prescription. We use $\alpha$ to define how efficiently 
the merger funnels the galaxy's gas to the
black hole accretion disk. To be consistent with our previous paper,
we adopt two definitions for a major merger: one with a mass ratio of
$M_{\rm p}/M_{\rm s} = 10$, and one with a mass ratio of 4. We adjust 
$\alpha_0$ to deactivate gas accretion below the major merger threshold. 
In Cox et al. 2008, the fitted $\alpha_0$ parameter suggests a nuclear 
starburst cut off for mass ratios larger than 9. Note that the 
Cox et al. 2008 study does not include the important effect of black hole 
feedback, which will shut off gas inflow after a merger.

Now that we have a more realistic description of the merger time for 
each black hole within a halo, we allow them to grow for a 
physically-motivated accretion timescale. The accretion of gas onto both 
incoming and central black hole starts when the two black holes are 
still widely separated, at the moment of the first 
pericenter passage, and continues until the black holes merge 
(c.f. Di Matteo et al. 2005, Colpi et al. 2007). This sets the 
accretion timescale, t$_{\rm acc}$, as follows: 
t$_{\rm acc}$ = t$_{\rm df}$(r=R$_{\rm vir}$) - t$_{\rm dyn}$(r=R$_{\rm vir}$),
where t$_{\rm df}$(r=R$_{\rm vir}$) is the merger timescale including 
dynamical friction; and t$_{\rm dyn}$(r=R$_{\rm vir}$) is dynamical time 
at virial radius R$_{\rm vir}$, which marks the first pericenter pass 
of the black hole. By stopping the accretion as the
black holes merge, we roughly model the effect of black hole feedback 
in stopping further accretion.

Putting these pieces together, the mass accreted by a black hole during 
t$_{\rm acc}$(r=R$_{\rm vir}$) is:

\begin{equation}
M_{\rm acc}=M_{\rm BH,0} (e^{\frac{\alpha t_{\rm acc}}{t_{\rm sal}}}-1),
\end{equation}

\noindent where M$_{\rm BH,0}$ is initial black hole mass, and $\alpha$ is 
starburst efficiency (Cox et al. 2008), and $t_{\rm sal}$ is defined above.
After t$_{\rm df}$, the incoming black hole merges with the SMBH at the 
center and a new SMBH is formed after having accreted gas for
 t$_{\rm acc}$. The accretion time and efficiency
both implicitly encode the large-scale dynamics of the merger and 
the bulk gas accretion into the nuclear region, while $t_{\rm sal}$ 
describes the accretion disk physics. As before, we set $t_{\rm sal}$
to describe sustained Eddington-limited accretion with a efficiency of
0.1 (Shakura $\&$ Syunyaev 1973).

\section{RESULTS}

Black hole mergers postponed for longer than the Hubble time will not occur, 
which reduces the merger rate over all redshifts. In the absence of 
dynamical friction, there are 1447 black hole mergers
in our volume between redshifts 20 and 0; this number reduces to 1248 
with Chandrasekhar dynamical friction, and 1063 with a Boylan-Kolchin estimate
for the dynamical friction decay time.
 
The number of dark matter halo mergers per unit redshift per year is 
presented in figure 3 (thick line). This halo merger rate directly 
corresponds to the black hole merger rate
when the black holes merge immediately after their host halos. When 
Chandrasekhar dynamical friction
is introduced to the black hole merger tree (figure 3, dashed line) 
the black hole merger rates decrease over all redshifts except at 
z $\leq$ 2.7, as the mergers that have been delayed from higher redshift 
begin to occur. The black hole merger rate is maximum at z=11 without 
dynamical friction, and at this redshift the rate decreases by a factor
of 1.5 for Chandrasekhar dynamical friction. This reduction in the black
hole merger rate is even more pronounced when we use the Boylan-Kolchin 
equation for the merger timescale (figure 3, dotted line), 
dropping the merger rate by a factor of two.

Those mergers that are postponed but have a merging timescale less 
than a Hubble time will be pushed down the merger tree toward lower 
redshifts. This is made explicit in figure 4, 
which plots the average change in redshift for a merger as a 
function of redshift if one includes dynamical friction. Black hole mergers
at redshift 10 are, on average, pushed to redshift 6, for example.
This results in an increase in the black hole merger
rate at low redshifts. We can see this clearly in figure 3; 
although the maximum merger rate at redshift z=11 has dropped, the number 
of mergers is larger than the Chandrasekhar merger rate at z $\leq$ 6 
(and therefore also larger than without dynamical friction at z $\leq$ 2.7). 
Again, this is because dynamical friction postpones mergers longer with 
the Boylan-Kolchin formula.  This suggests higher LISA merger rates at 
low redshifts than any estimate that used the Chandrasekhar formula. 
In addition, since each merger occurs at a lower redshift, it will 
certainly be a louder gravitational wave source, and any associated
electromagnetic signature will be brighter, as well. We will return 
to the LISA detectability in a future paper.

Interestingly, including dynamical friction from gas at distances smaller 
than 1 kpc does not make difference in our merger rates. Those black 
holes that reach 1 kpc from the galactic center will reach a parsec even 
without the gas -- the gas timescale is simply invoked to ensure that 
the black holes pass through the 3-body scattering stage efficiently. 
For the rest of this paper, we adopt the Boylan-Kolchin dynamical friction 
fit as a realistic treatment of the black hole merger timescale.

In order to address the effect of inefficient black hole seed formation 
on the black hole merger rates, we randomly seed only $10\%$ of those 
halos that could host a Pop III star with a $200 \Msun$ black hole. 
We ran 100 realizations of this process, applying Boylan-Kolchin merger 
timescales as described above. These rates are presented at the 
bottom of figure 3 with maximum of range 2 - 3 yr$^{-1}$. This is a 
pessimistic scenario since it suggests that only 10$\%$ of dark matter 
halos are capable of hosting a seed black hole. Future work on this 
subject will give more realistic predictions for Pop III seeding
efficiency, but we point out that even with our lowest estimate for 
the number of black hole seeds, the predicted LISA merger rate of 
2 - 3 yr$^{-1}$ makes this population a reliable gravitational wave source.

By design of our small volume simulation, we have only one SMBH in our 
volume sitting a halo with $3 \times 10^{12} \Msun$, roughly the mass of 
the Milky Way halo. Figure 5 shows the SMBH mass at the center of the 
primary halo as a function of redshift.  SMBH growth without dynamical 
friction is presented with thin lines for the three scenarios 
(dry growth, 4:1 growth, and 10:1 growth) described previously. The thick 
lines present corresponding growth models with dynamical friction. The 
first merger of the primary dark matter halo with a satellite occurs 
at redshift z=14.2; without dynamical friction, that is also the
redshift of the first black hole merger. This merger is postponed until 
z=11.7 when dynamical friction is added. In our model, major mergers are 
followed by the episodes of gas accretion which can be identified in 
figure 5 as rapid increases in the SMBH mass. A great example
of how SMBH mergers are postponed by dynamical friction can be seen 
in the change in SMBH mass for the redshift range 4.7 $\leq$ z $\leq$ 6.1. 
Although the last major merger of the primary halo is at z=6.1 (figure 5, 
thin dash-dot line), the SMBH at the center of the primary halo undergoes 
at least three rapid increases in mass in the 10:1 growth model 
(figure 5, thick dash-dot line) for the redshift range 
4.7 $\leq$ z $\leq$ 6.1. These are from satellite mergers originating at 
higher redshift.  Redshift 4.7 marks the end of the gas accretion era 
for the SMBH, well after it experiences its last major merger.
The SMBH is close to its final mass at this time, as the only channel 
left in our model for increasing its mass is via black hole mergers. We 
note that the 10 next most massive black holes all reach their final 
mass at varying redshifts between 10 and 1, though in our volume, these 
black holes are $10^3-10^5 \Msun$.

The range for the SMBH mass set by the 4:1 and 10:1 dynamical friction 
growth models is 1.7$\times 10^6\Msun$ $\leq$ M$_{\rm SMBH}$ $\leq$ 
1.3$\times 10^7\Msun$. If we neglect dynamical friction, the black hole masses 
are unphysically large. Dynamical friction removes 27$\%$ of the total 
number of initially merging black holes because these will take 
longer than a Hubble time to coalesce. All other black holes merge 
later, and these effects both prevent the seed from accumulating a large 
mass early on to use as a 'base' for Salpeter accretion phases. The effect
of delaying and removing the mergers outweighs the longer accretion
timescales seen when dynamical friction is included.

Figure 6 shows the universal merger rates for those dark matter halos that 
merge in less than a  Hubble time within the mass range 
10$^7\Msun$ $\leq$ M$_{\rm DMH}$ $\leq$ 10$^{12}\Msun$, for various mass 
ratios and combined masses of merging dark matter halos. Black holes seeded 
in halos in the middle and bottom panels grow through
mergers only and never accrete, while gas accretion impacts black hole 
growth in 4:1 and 10:1 halo mergers (figure 6, upper panel). In our volume, 
all the mergers that occur at redshifts z $\leq$ 2 will not finish in less 
than a Hubble time, as these mergers are all high mass ratio. We remind the 
reader, though, that these halo merger rates are extrapolated from
our 10 Mpc simulation box; a better global rate for these low mass halos 
will be achieved with more of these volumes. 

Since the merger timescale is larger for higher mass ratio halos,
the black hole gas accretion time scale is longer, as well. Figure 7 
shows the gas accretion time scale for black holes hosted by dark matter 
halos that merge with mass ratio M$_{\rm DMH1}$/M$_{\rm DMH2}$ and combined 
mass p=log(M$_{\rm DMH1}$+M$_{\rm DMH2}$). Notice that gas accretion is 
activated only for a small range of halo mass ratios 
M$_{\rm DMH1}$/M$_{\rm DMH2}$ $\leq$ 10. For these mergers, the black holes 
do not accrete for longer than $\sim$ 1000 Myr, and the accretion 
efficiency is strongly damped as the halo mass ratio tends toward 10:1.

The number of accreting black holes as a function of redshift is presented 
in figure 8. The thick line represents the 10:1 growth case, and this 
also corresponds to halos with mass ratio 
M$_{\rm DMH1}$/M$_{\rm DMH2}$ $\leq$ 10 in figure 7. In the same manner, 
the thin line (4:1 growth) corresponds to halos with mass ratio 
M$_{\rm DMH1}$/M$_{\rm DMH2}$ $\leq$ 4 in figure 7. 
In either accretion scenario, the number of accreting black holes reduces
by $\sim$ 200 at redshift 4.7, from a few hundred accreting black holes 
to a few tens. This corresponds to the redshift at which the SMBH at the 
center of the primary halo stops accreting gas from the previous major merger. 
Those black holes that are still accreting gas at z $\leq$ 4.7 are mostly 
rogue black holes at the outskirts of the primary halo. We will discuss 
these more in the next subsection.

The last episode of gas accretion by the SMBH is also an epoch 
during which the SMBH accumulates largest amount of mass. Figure 9 shows 
the rate of mass growth of the SMBH for 10:1 growth with (thick line) and
without (thin line) dynamical friction. At the redshift of the 
last gas accretion, z $\sim$ 4.7, the mass growth rate 
is $\sim$ 1 $\Msun$/year for 10:1 growth with dynamical friction. 
Thereafter, the gas accretion is extremely damped; although this SMBH is
far too low mass to be considered a high luminosity AGN analogue, 
it may be appropriate to link this SMBH with a low luminosity AGN (Ho 2008).

We divide the universal black hole merger rates as a function of redshift
in figure 10 into different binary mass ratios and total binary mass ranges, 
where p = log (m$_{\rm 1}$ + m$_{\rm 2}$) is a measure of the total binary 
black hole mass. Horizontally, we separate the merger rate by mass ratio: 
in panels a, b, and c, we focus on 1$\ltsim$m$_{\rm 1}$/m$_{\rm 2}<10$;
10$\ltsim$m$_{\rm 1}$/m$_{\rm 2}<100$ in d, e, f; and 
100$\ltsim$m$_{\rm 1}$/m$_{\rm 2}<10000$ in g, h, i. Each column represents 
a different black hole growth scenario: we cover dry growth in a, d, g; 
4:1 growth in b, e, h; and 10:1 growth in c, f, i. 
The  black hole mass ratio of 100$\ltsim$m$_{\rm 1}$/m$_{\rm 2}<10000$
corresponds to SMBH/IMBH mergers, and  
the merger rate is 3 - 7 yr$^{-1}$ for both the 4:1 case (h) and 10:1 case 
(i), between redshifts 1 and 5. Assuming that before
the merger occurs, the incoming IMBH behaves as ULX source, the expected 
frequency of ULX sources would also be $\sim$ 3 - 7 yr$^{-1}$. These 
black hole mergers have the appropriate total mass to be important LISA 
sources. We will return to the integrated LISA signal, event rates, and 
source characterization in a future paper.

\subsection{Rogue Massive Black Holes}

Over the span of the simulation, the primary accretes 483 satellite 
halos total, and 382 of them have orbital decay times that are longer 
than a Hubble time (see figure 11 for the merger rate 
and mass ratios of these halos). By redshift zero, then, these halos 
are still orbiting at significant distances from the primary center -- 
and although the halos themselves have been stripped by the primary 
potential, they are still embedded with massive black holes.
Figure 12 displays the number of these 'rogue' black 
holes as a function of the distance from the primary center at redshift 
zero. For a primary scaled to the Milky Way, this implies that roughly 
1-10 of these massive black holes in orbit at the solar radius.

Given the merger-driven accretion scenario described in the previous section,
we can determine the mass of these black holes at redshift zero. 
Figure 13 shows the mass of each rogue as a function of
distance from the primary center; each black hole is color-coded to
keep track of when the satellite halo entered the virial radius of the 
primary. Note that the most massive of these black holes are embedded in 
satellites that have themselves undergone mergers before entering the
primary halo at redshift $2<z<4$. This 'preprocessing' via prior mergers allows
the black holes to enter the primary halo at later times with higher masses;
these most massive rogues remain at roughly half the virial radius by
redshift zero.

There is sufficient evidence that the halos of galaxies, from massive 
ellipticals to isolated spirals, can host diffuse hot halo gas 
(Matthews $\&$ Brighenti 2003, O'Sullivan 2001, Pedersen et al. 2006). 
As these black holes orbit within the primary halo, they can
accrete from this ambient hot halo gas via Bondi-Hoyle accretion. Here,
the mass accretion rate can then be described by:

\begin{equation}
{{\dot M_{\rm BH}}} = {{4 \pi G^2 M_{\rm BH}^2 \rho_{\rm ISM}} \over { (c_s^2 + v^2)^{3/2}}},
\label{eqn:Bondi}
\end{equation}

\noindent where $\rho_{\rm ISM}$ is the density of the halo gas, $c_s$ is the
gas sound speed, and $v$ is the velocity of the black hole. For these rogue 
black holes, we add this more quiescent form of accretion from the time the
black hole has entered the virial radius of the halo in order to determine
whether any of these rogue black holes may be visible today. To model the hot 
halo gas, we assume it has an isothermal density profile consistent with X-ray 
observations of halo gas in ellipticals (Matthews $\&$ Brighenti 2003), and that 
the ideal gas is in hydrostatic equilibrium with the gravitational potential 
of the halo. We assume a gas core radius of $0.3 R_s$, where $R_s$ is 
the scale radius of the NFW halo, and that the gas temperature is the 
virial temperature at the virial radius. This yields a central gas 
temperature of $\sim 2 \times 10^6 K$ for a dark matter halo of mass 
$2 \times 10^{12} M_\odot$, which is also consistent with 
observations (Petersen et al. 2006). At redshift zero, we assume 
the black holes are on bound elliptical orbits of eccentricity $0.8$ to 
match simulation predictions for satellite mergers (e.g. Ghigna et al. 1998, 
Sales et al. 2007). 

By redshift zero, a black hole will have grown to a new mass 
$M_{\rm rogue}$, and will have a new accretion rate from 
equation~\ref{eqn:Bondi}. The radiated luminosity from this process 
can be determined from $L_{\rm rogue} = \eta {\dot M_{\rm BH}} c^2$, 
where $\eta$ describes how efficiently mass is converted to energy;
the standard assumption for an accretion disk around a spinning black 
hole sets $\eta\sim 0.1$ (Shakura $\&$ Syunyaev 1973). We find that
our most massive rogue black holes can radiate at $\sim 10^4 L_\odot$ (figure 14), 
making them somewhat fainter than the brightest ULXs; the fact that 
the hot halo gas is so tenuous at these radii is what accounts for the 
relatively low luminosity. However, if these rogue black holes are common, 
they will be observable in the outskirts of the Milky Way halo with Chandra, 
and the most luminous can have sufficient signal-to-nose for spectral 
characterization. Note, though, that a significant fraction of these may 
have advection-dominated accretion flows, rather than proper accretion disks, 
and hence $\eta$ may be much less than 0.1; this would naturally make the 
black hole less luminous, but with a harder spectral signature. Even if they 
are not observable electromagnetically, they may be detected by mesolensing
(di Stefano 2007).

\subsection{Low Mass Black Hole Demography}

For dark matter halo masses above roughly $5 \times 10^{11} \Msun$,
the dark matter halo mass correlates well with the mass of the 
supermassive black hole (Ferrarese 2002):

\begin{equation}
{M_{\rm BH}\over {10^8 \Msun}} \sim 0.10 \Bigg( {{M_{\rm DMH}} \over {10^{12} \Msun}} \Bigg)^{1.65},
\end{equation}
where $M_{\rm DMH}$ is the dark matter halo mass. Below this mass, there is tentative 
evidence that halos are less effective at forming massive black holes, 
and may even be unable to form them (Ferrarese 2002). To be fair, though, 
if this relation does hold down to a $10^{11} \Msun$ halo, the expected
$2 \times 10^4 \Msun$ black hole would be difficult to detect, 
observationally. 

In our volume, there are few halos with masses $\geq 10^{11}\Msun$ by 
redshift zero, so our sample is limited. However, we can still find the final
black hole and halo mass for every black hole that has
 undergone some accretion by redshift zero. This is plotted in figure 15.
Notice that our single SMBH in the volume is consistent with the 
$M_{\rm BH} - M_{\rm DM}$ relation for a 10:1 growth scenario, and many
of our lower mass black holes do as well. There are a few tidally-stripped 
satellite halos on the low mass end that host overweight black holes. 
In addition, those dark matter halos with sparse merger histories host 
underweight black holes. This may point to a dependence of the black hole 
mass on environment. We will be able to explore this more carefully with 
the set of simulations currently underway.

\subsection{Comparison to Published Black Hole Merger Rates}

Previous work on black hole merger rates use semi-analytical models
based on Extended Press-Schechter theory (EPS), and the dynamical friction 
in these models uses the Chandrasekhar formalism. We compare our results 
with those obtained from EPS theory. Figure 16 shows the 
merger rates for four Press-Schechter models described in Sesana et al. 2007.
In the VHM model, massive DMHs (M$_{\rm DMH}$=10$^{11}$ - 10$^{15}$ $\Msun$)
are seeded with m$_{\rm BH}$$\sim$200$\Msun$ black holes at z=20; in the 
KBD model,low mass halos (M$_{\rm DMH}$=10$^6$ - 10$^7$ $\Msun$) are 
seeded with m$_{\rm BH}$$\sim$5$\times10^4\Msun$ at 15$\ltsim$z$\ltsim$20; 
and the BVR models explore different redshift ranges for seeding black holes 
within halos: m$_{\rm BH}$=$10^4 - 10^5 \Msun$
at 15$\ltsim$z$\ltsim$20 in the BVRhf model and 18$\ltsim$z$\ltsim$20
in the BVRlf model. Our black hole merger rate is overplotted in thick black. 
The best match to our model is the KBD model, though their seed black holes 
are more massive. This match is not surprising since the range of seeding 
redshifts and the halo mass range are similar to ours. Our merger rate at 
high redshift is larger because we seed black holes in a wider mass range 
of dark matter halos. At z $\leq$ 11, however, KBD merger rates are 
larger since they use Chandrasekhar dynamical friction, which 
underestimates the merger timescale. This simple rate comparison of 
merger rates, however,  neglects the larger differences in black hole 
masses and the mass ratio of each merger.

\section{DISCUSSION}

Using a small volume cosmological n-body simulation to construct the 
merger history of low mass dark matter halos, we have studied the growth 
and merger rate of black holes from seeds at redshift 20 to low mass SMBHs 
at redshift zero, analytically incorporating the important sub-resolution 
physics to model the black hole dynamics and gas accretion.
We found that the method used to estimate the dynamical friction 
timescale makes a large difference in the total rate and redshift 
distribution of black hole mergers; for example, the merger rate is smaller 
at higher redshift z $\geq$ 6 with an n-body based dynamical friction 
estimate (Boylan-Kolchin 2008) than with the Chandrasekhar dynamical 
friction approximation, because this approximation underestimates the 
true merger time. If most of the black hole growth is tied to gas 
accretion that is activated by major mergers, then using the Chandrasekhar 
dynamical friction approximation will overestimate the mass 
of the black hole. Moreover, these larger, more realistic merger 
timescales will postpone black hole mergers to lower redshifts, 
which increases the black hole merger rate at z $\leq$ 6. 

We find that the maximum black hole merger rate is 
R $\sim 25 \rm yr^{-1}$ at z $\sim 10 \rm yr^{-1}$ .
However, when a Pop III seed formation efficiency of 0.1 is adopted, 
the maximum merger rate falls to R $\sim 3 \rm yr^{-1}$ . This scenario 
is based on the most pessimistic picture of the evolution of the first stars; 
future work will constrain the occupation fraction of seed black holes. 
However, even with such a low Pop III formation efficiency, and a 
larger merger timescale than has previously been used, the merger 
rate is large enough for SMBH mergers to remain an important LISA source.

We also introduce an expression for the growth of a black hole 
that includes direct mergers of the black hole and Eddington-limited 
accretion driven by major mergers. We use the merger time starting from the 
first pericenter passage of the incoming black hole to set the gas 
accretion timescale, and we weighted the efficiency of accretion 
by the available gas in the galaxy nucleus, which has been 
carefully studied for unequal mass galaxy mergers (Cox et al. 2008).
With this black hole growth scenario, we find that if we define a major 
merger as one with a $4:1$ halo mass ratio, the final SMBH
mass at the center of a Milky Way type halo is
M$_{\rm BH}$ = 1.7$\times 10^6\Msun$. If, instead, we adopt a more
lenient definition of a major merger (10:1),  
$ M_{\rm BH} $ = 1.3$\times 10^7\Msun$. This first value is similar 
to the mass of the SMBH observed at the center of our Galaxy, while the
second is consistent with the mass at the center of M31. Hence, we 
have shown in this study that not only is it possible to grow these 
low mass SMBHs from very light Pop III remnants, but that most of the 
SMBH mass can be accounted for with a combination of major merger-driven 
gas accretion and direct black hole mergers.

We emphasize that with a growth prescription that is so strongly tied 
to major galaxy mergers, it is possible to have low mass black holes 
in place at high redshift and not accreting strongly thereafter.
In fact, the most massive black hole in our small volume grew most 
rapidly for a short span around redshift 4.5; thereafter, it was quiescent. 
This may seem in direct contradiction to previous observational studies 
that have argued for the ``anti-hierarchical'' growth of black holes,
meaning that the most massive SMBHs accrete most at high redshift, 
while the lower mass SMBHs are still accreting (and grow the most) 
at $z<1$ (Merloni 2004, Marconi et al. 2004). This apparent
contradiction may not be so clear, however. First, the black holes
in our volume were lower mass than what is typically considered in 
the local census of relic AGN. Second, these observational studies 
are based on linking the observed local black hole mass function to 
high redshift AGN assuming that these black holes grow only by gas 
accretion (and not merging) with a single accretion efficiency. Our 
black holes would not be able to grow to $10^{6} \Msun$ without a 
rich merger history. Third, our simulations do not explicitly 
include any of the more 'quiescent' black hole growth modes, such as
bar-driven gas inflow, stellar tidal disruption, compact object capture, 
and sub-Eddington Bondi-Hoyle-Littleton accretion from the surrounding 
gaseous medium. These may very well act to increase the growth rate 
of some of the black holes in our simulation at lower redshift{\footnote{They 
also may have little effect: a recent study by Ho, Filippenko 
$\&$ Sargent (1997) found that there barred galaxies do not increase 
the strength or frequency of the AGN}}. However, if this form of growth 
were to dominate at $z<1$ for the SMBH in our volume, the SMBH would be 
too massive.  Finally, we appeal to cosmic variance; our volume was
selected to be a sparse group environment, and as such, SMBHs were rare. 
If galaxy merger-driven inflow, direct black hole mergers, and these more 
quiescent forms of black hole accretion all play roles in the assembly of 
low mass SMBHs, each growth channel may have a different strength
as a function of environment. We are running several high-resolution, 
small volume simulations to tackle this issue.

In this paper, we have neglected gravitational wave recoil, a 
potentially important mechanism that may inhibit black hole growth.
Binary black holes strongly radiate linear
momentum in the form of gravitational waves during the plunge
phase of the inspiral -- resulting in a ``kick'' to the new black hole. 
This, in itself, has long been predicted as a consequence
of an asymmetry in the binary orbit or spin configuration. 
Previous kick velocity estimates, though, were either highly uncertain 
or suggested that the resulting gravitational  wave recoil velocity was 
relatively small, astrophysically speaking. Now, recent 
results indicate the recoil can drive a gravitational wave kick velocity as 
fast as $\sim$4000$\kms$ (Herrmann et al. 2007; 
Gonzalez et al. 2007a, 2007b; Koppitz et al. 2007; Campanelli et al. 2007, 
Schnittman $\&$ Buonanno 2007). In reality, much smaller values than 
this maximum may be expected in gas-rich galaxies due 
to the alignment of the orbital angular momentum
and spins of both black holes (Bogdanovic et al. 2007). 
However, even typical kick velocities ($\sim 200~\kms$) are interestingly 
large when compared to the escape velocity of typical astronomical systems - 
low mass galaxies, as an example, have an escape velocity of $\sim 200~\kms$ (e.g. Holley-Bockelmann et al. 2007). 
The effect of large kicks combined with low escape velocity from the 
centers of small dark matter halos at high redshift plays a major role in
suppressing the growth of black hole seeds into SMBH.
Even the most massive dark matter halo at 
z$\geq$11 can not retain a black hole that receives 
$\geq$ 150 ${\rm km \,s^{-1}}$ kick (Merritt et al. 2004, Micic et al. 2006). 
In our next paper we will incorporate the effect of recoil velocity on
the expected merger rates and the growth of the SMBH.

\section*{ACKNOWLEDGMENTS}

\clearpage

\begin{figure}
\vspace{0.5in}
\begin{center}
\includegraphics [width=4.in,angle=0]{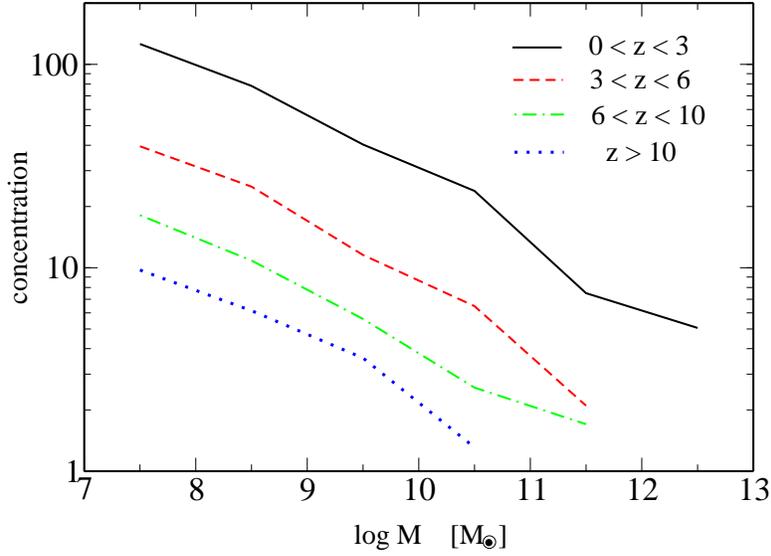}
\caption[Fig 1.]{Average concentration parameter in four redshift 
ranges for all dark matter halos in our simulation as a function of halo mass. }
\end{center}
\end{figure}

\begin{figure}
\vspace{0.5in}
\begin{center}
\includegraphics [width=4.in,angle=0]{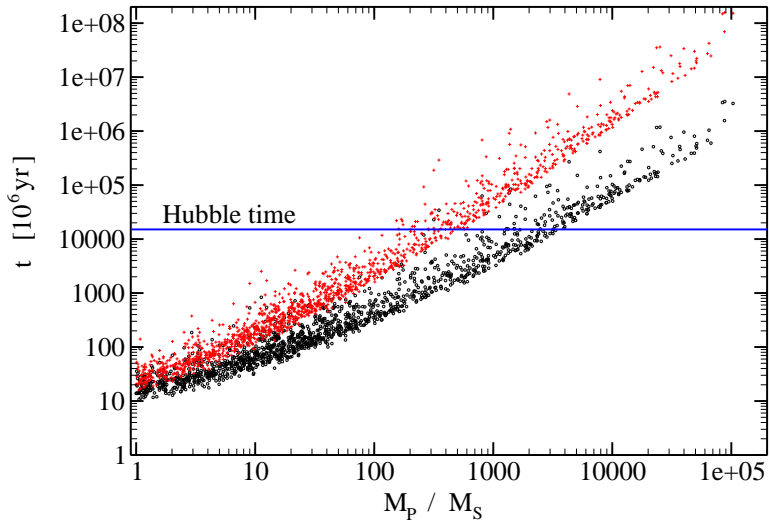}
\caption[Fig 2.]{Time for each satellite to reach the primary halo center 
as a function of the mass ratio of merging halos. Black circles represent the
Chandrasekhar dynamical friction time and red pluses represent the merging time 
calculated from a simulation-based numerical fit (Boylan-Kolchin 2008). 
Both timescales are compared to the Hubble time. Merging halos above the 
horizontal blue line will not finish their merger and are removed from the 
merger tree, but their positions are updated within the primary halo.}
\end{center}
\end{figure}

\clearpage

\begin{figure}
\vspace{0.5in}
\begin{center}
\includegraphics [width=4.in,angle=0]{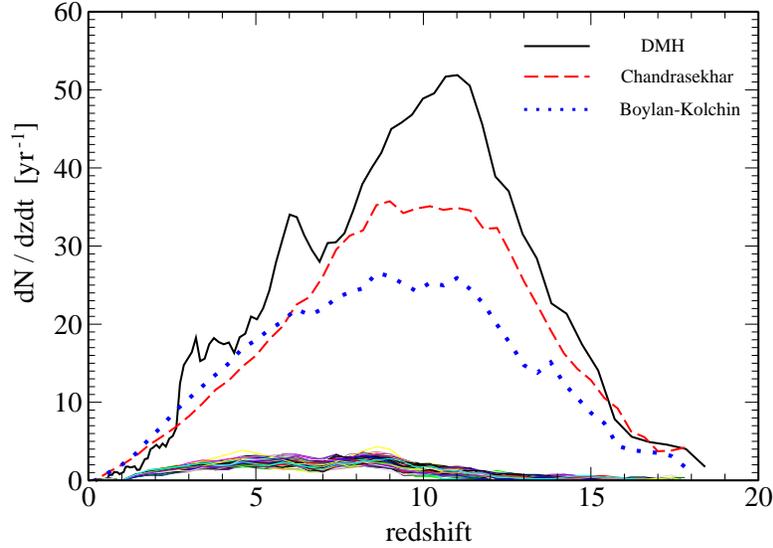}
\caption[Fig 3.]{Top line (black):  Black hole merger rates without dynamical 
friction. Note that in this case, this is equivalent to  the dark matter halo 
merger rate; Dashed (red): Black hole merger rate with the Chandrasekhar dynamical 
friction formula; Dotted (blue): Black hole merger rate obtained from the Boylan-Kolchin 
numerical fit for dynamical friction. We predict that LISA black hole merger rates
at z $\leq$ 2.7 are larger than estimated by models based on Chandrasekhar dynamical 
friction. Bottom: 100 realizations of black hole merger rates for an initial seed black 
hole occupation fraction of 0.1, with Boylan-Kolchin dynamical friction. The maximum 
black hole merger rate of 3 is considered a pessimistic scenario.}
\end{center}
\end{figure}

\begin{figure}
\vspace{0.5in}
\begin{center}
\includegraphics [width=4.in,angle=0]{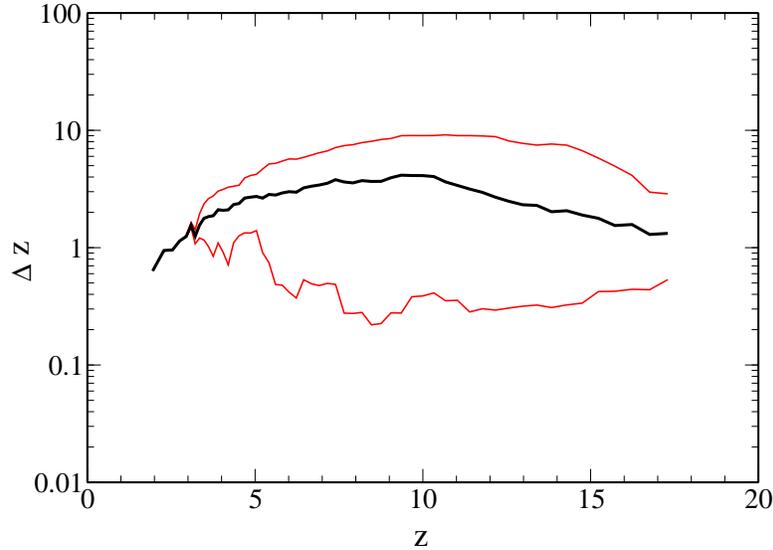}
\caption[Fig 4.]{Thick black - average change in the merger redshift when dynamical 
friction is applied as a function of merger redshift (without dynamical friction). 
Red lines represent minimum and maximum shift in the redshift of black hole merger. 
Dynamical friction postpones black hole mergers toward lower redshifts, making them 
louder LISA sources and increasing the local rate as well.}
\end{center}
\end{figure}

\begin{figure}
\vspace{0.5in}
\begin{center}
\includegraphics [width=4.in,angle=0]{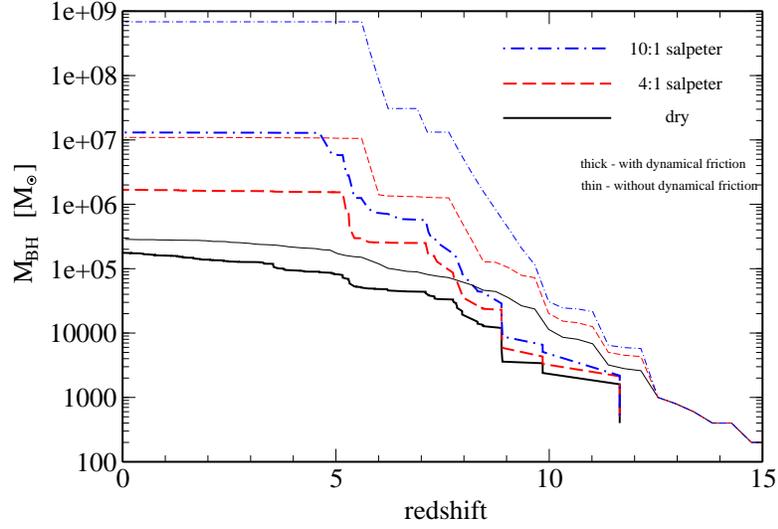}
\caption[Fig 5.]{Mass of the central SMBH as a function of redshift. Thin lines 
correspond to SMBH growth using the prescription described in the text when dynamical 
friction is neglected. Thick lines represent SMBH growth with dynamical friction. 
``Dry'' growth is in black; ``4:1'' growth in red; and ``10:1'' growth in blue. 
The final episode of gas accretion onto the SMBH is at redshift z=4.7 in the 10:1 
case with dynamical friction. Later on, the SMBH grows through mergers 
only and its mass remains almost constant until z=0.}
\end{center}
\end{figure}

\begin{figure}
\vspace{0.5in}
\begin{center}
\includegraphics [width=4.in,angle=0]{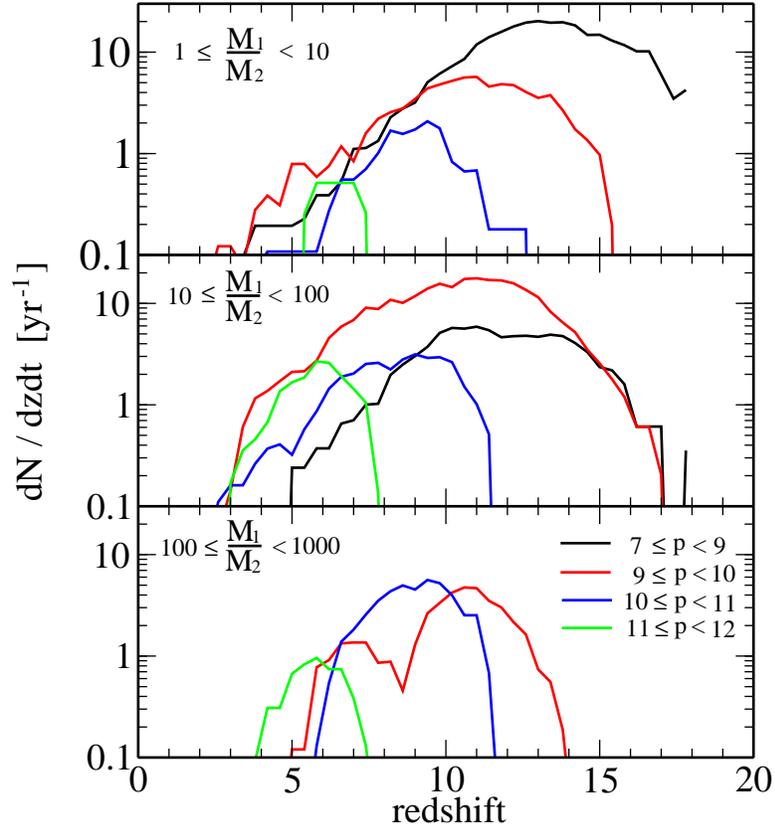}
\caption[Fig 6.]{Dark matter halo merger rates as a function of redshift for all halos
that finish merging by redshift zero. Three panels show various ranges for the halo 
mass ratio and the total combined halo mass. M$_1$ is mass of the primary halo, 
M$_2$ is mass of the satellite, and the total mass is defined by p = log (M$_1$ + M$_2$). 
The panel with M$_1$/M$_2$ $\leq$ 10 shows merger rates for major mergers that activate 
gas accretion onto their black holes.}
\end{center}
\end{figure}

\begin{figure}
\vspace{0.5in}
\begin{center}
\includegraphics [width=4.in,angle=0]{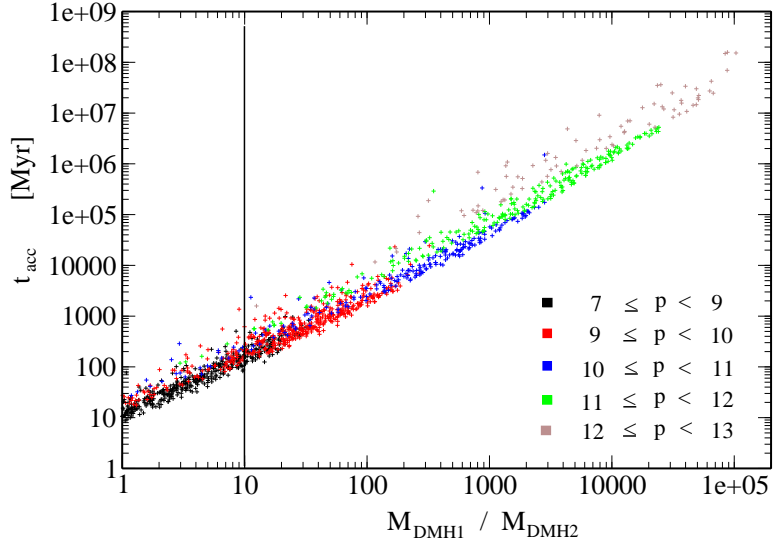}
\caption[Fig 7.]{Gas accretion time, $t_{\rm acc}$, for black holes at the center of a 
satellite halo of mass M$_{\rm DMH1}$ that is merging with a primary halo of mass 
M$_{\rm DMH2}$. The total mass is denoted by p = log (M$_{\rm DMH1}$ +M$_{\rm DMH2}$), and 
t$_{\rm acc}$ = t$_{\rm merger}$ - t$_{\rm dyn}$ where
t$_{\rm merger}$ is the merger time scale, and t$_{\rm dyn}$ is 
the dynamical time,  making accretion begin at the first pericenter pass. 
 Only merging halos with mass ratio less than 10
(vertical line) will allow gas accretion onto the black holes.}
\end{center}
\end{figure}

\begin{figure}
\vspace{0.5in}
\begin{center}
\includegraphics [width=4.in,angle=0]{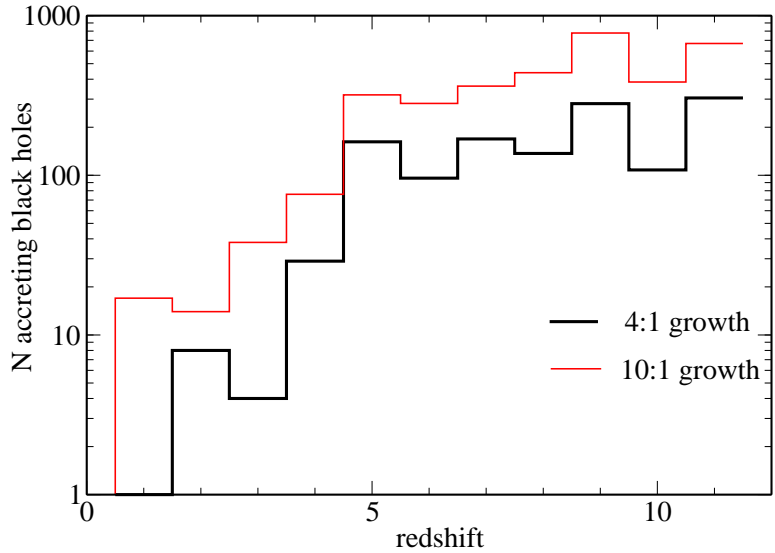}
\caption[Fig 8.]{Number of accreting black holes as a function of redshift. Here we 
include both black holes that do and do not eventually merge. The rapid drop in the number
of accreting black holes (for both 4:1 and 10:1 Salpeter growth) at redshift z $\sim$ 5 corresponds 
to the end of SMBH growth at the primary halo center (Figure 4). The remaining black holes at low
redshift correspond to black holes that are still sinking into the center by redshift zero.}
\end{center}
\end{figure}

\begin{figure}
\vspace{0.5in}
\begin{center}
\includegraphics [width=4.in,angle=0]{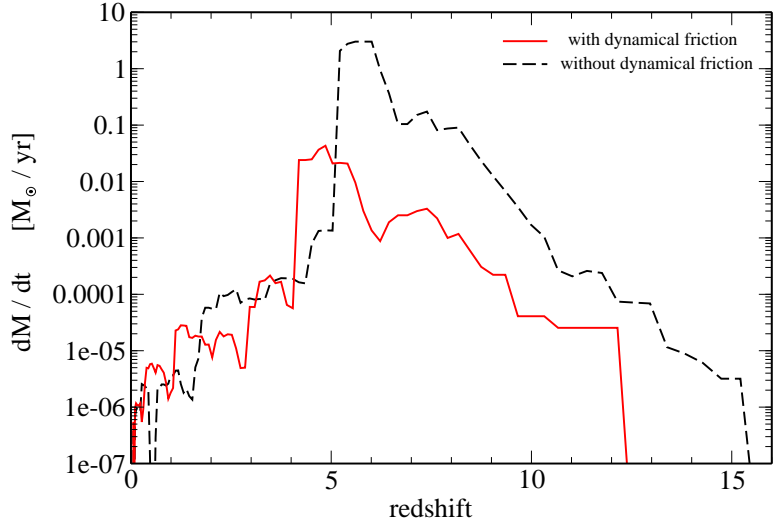}
\caption[Fig 9.]{Mass growth rate of the SMBH as a function of redshift. Black dashed: ``10:1'' growth
without dynamical friction. Red thick: ``10:1'' growth with dynamical friction. The SMBH growth is 
largely complete by redshift 4.7, as the merger-driven gas supply is depleted.}
\end{center}
\end{figure}


\begin{figure}
\vspace{0.5in}
\begin{center}
\includegraphics [width=4.in,angle=0]{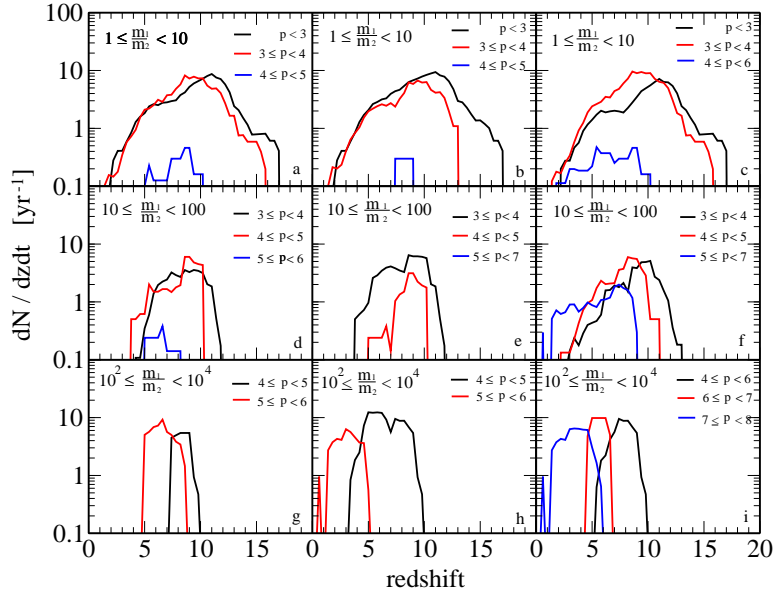}
\caption[Fig 10.]{Black hole merger rates as a function of redshift
for different binary mass ratios and total binary mass ranges,
p = log (m$_{\rm 1}$ + m$_{\rm 2}$). Horizontally, we separate the 
merger rate by mass ratio: in panels a, b, and c, 
we focus on 1$\ltsim$m$_{\rm 1}$/m$_{\rm 2}<10$;
10$\ltsim$m$_{\rm 1}$/m$_{\rm 2}<100$ in d, e, f; and 100$\ltsim$m$_{\rm 1}$/m$_{\rm 2}<10000$
in g, h, i. Each column represents a different black hole growth scenario:
we cover dry growth in a, d, g; 4:1 growth in b, e, h; and 10:1 growth in c, f, i. 
For redshifts z $\sim$ 1 - 5 and mass ratios of 100$\ltsim$m$_{\rm 1}$/m$_{\rm 2}<10000$, 
the merger rate is 3 - 7 yr$^{-1}$ in 4:1 case (h) and 10:1 case (i). These black hole 
mergers are in the appropriate mass and redshift range to be important LISA sources.}
\end{center}
\end{figure}

\begin{figure}
\vspace{0.5in}
\begin{center}
\includegraphics [width=4.in,angle=0]{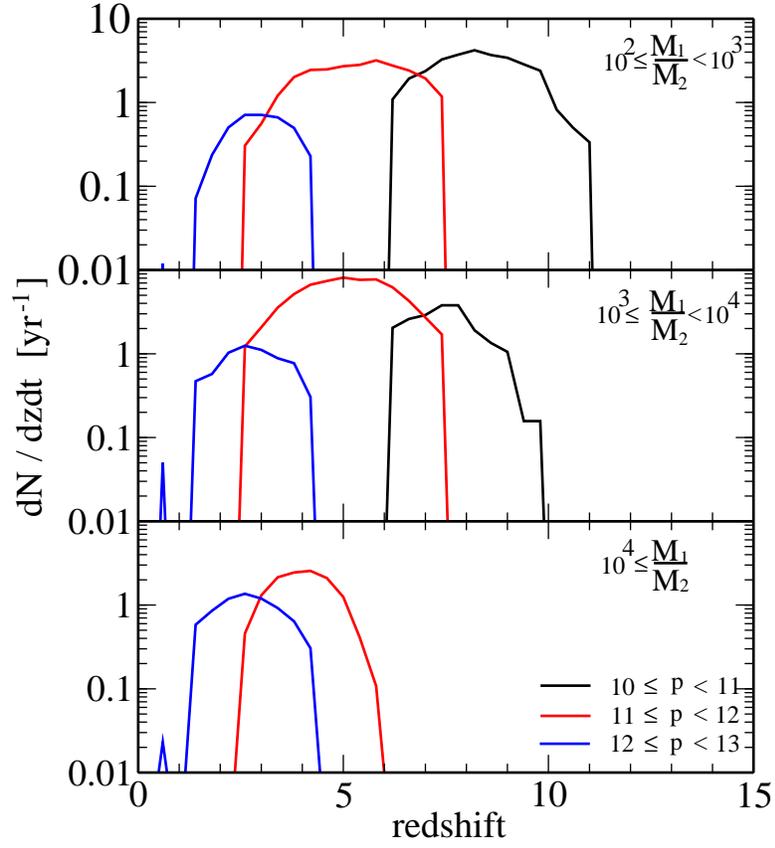}
\caption[Fig 11.]{Rate that the primary captures rogue black holes. Most of the seed 
black holes were captured before redshift 5 and were hosted by low mass halos.}
\label{fig:roguehalo}
\end{center}
\end{figure}

\begin{figure}
\vspace{0.5in}
\begin{center}
\includegraphics [width=4.in,angle=0]{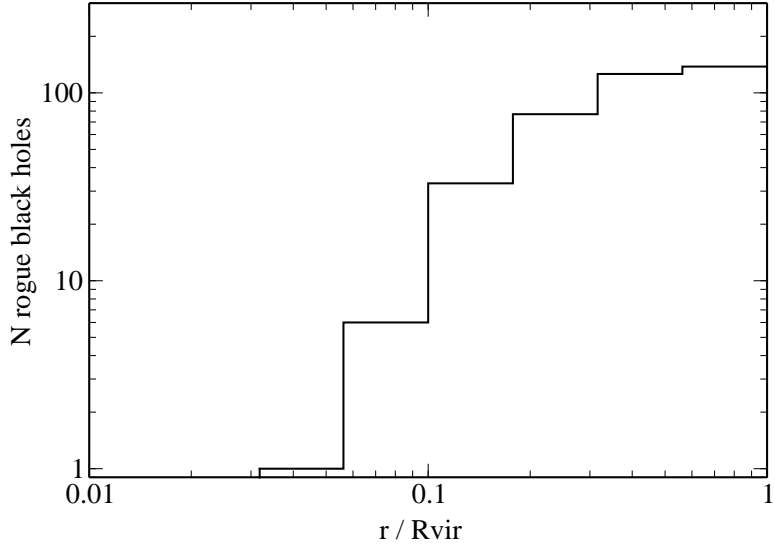}
\caption[Fig 12.]{Distribution of massive rogue black holes inside the primary halo
at redshift zero. The dark matter hosts of these rogue black holes have merged with 
the primary halo but the dynamical friction time is longer than the Hubble time. Hence,
these massive black holes do not sink to the center by z=0.}
\label{fig:roguedist}
\end{center}
\end{figure}

\begin{figure}
\vspace{0.5in}
\begin{center}
\includegraphics [width=4.in,angle=0]{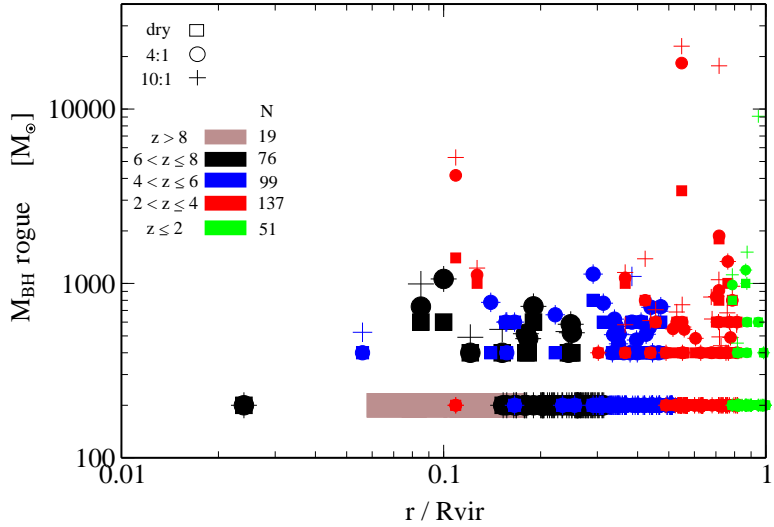}
\caption[Fig 13.]{Rogue black hole mass as a function of virial radius for three
accretion models with dynamical friction: ``dry'' in squares; ``4:1'' in
circles; and ``10:1'' in pluses. Colors correspond to the redshift range where black hole
became rogue -- this is also the merger redshift of merger the primary and satellite halos. 
Many of the most massive rogue black holes are accreted late and have been 'pre-processed' 
with previous mergers of their own.}
\label{fig:roguemass}
\end{center}
\end{figure}

\begin{figure}
\vspace{0.5in}
\begin{center}
\includegraphics [width=4.in,angle=0]{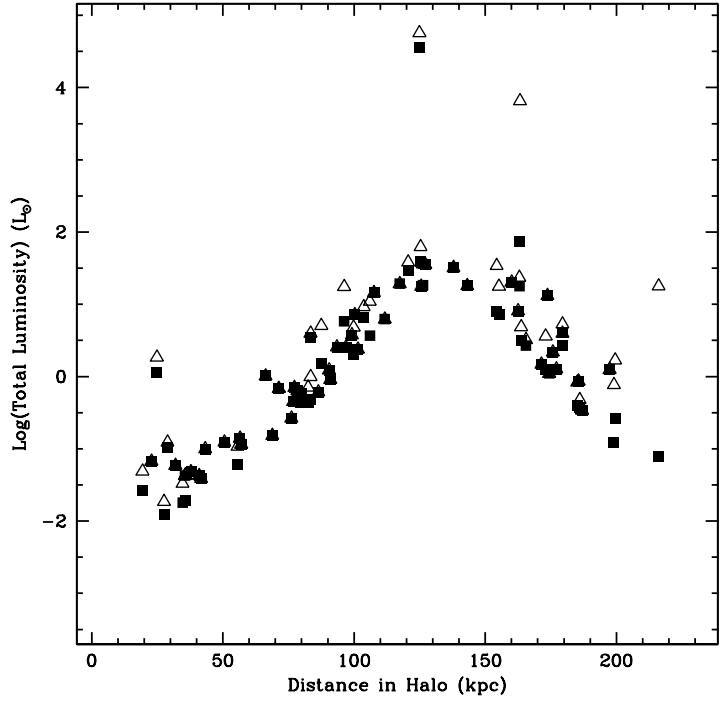}
\caption[Fig 14.]{Bolometric luminosity of rogue black holes as a function of distance 
from the primary halo center. Black holes are assumed to accrete via a Bondi-Hoyle mechanism 
from the ambient gas in the primary. Blue squares represent out 10:1 accretion scenario, 
and red squares represent the 4:1 scenario. Most black holes will have roughly solar 
luminosities, but a few are expected to be luminous Xray sources (just under the ULX cut-off) 
that reside in the Milky Way halo.}
\label{fig:rogueulx}
\end{center}
\end{figure}

\begin{figure}
\vspace{0.5in}
\begin{center}
\includegraphics [width=4.in,angle=0]{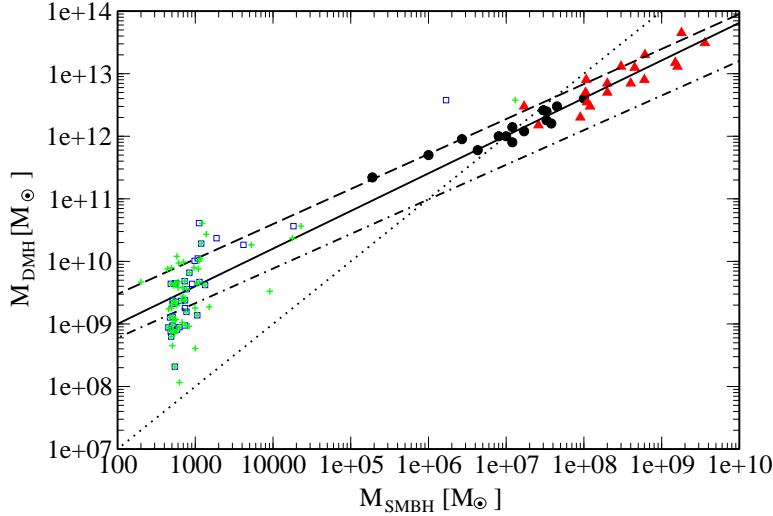}
\caption[Fig 15.]{Dark matter halo mass as a function of central SMBH mass for 
spiral galaxies (filled circles) and elliptical galaxies (filled triangles) 
(data taken from Ferrarese 2002). The dotted line represents M$_{\rm DMH}$ / M$_{\rm BH}$ = 10$^5$. 
The solid line obtains a halo mass using Bullock et al. 2001 prescription to
relate the virial velocity, v$_{\rm vir}$, to the circular velocity, v$_{\rm c}$. 
The dashed line shows the same function when v$_{\rm vir}$ = v$_{\rm c}$, and
the dot dashed line is for v$_{\rm c}$ / v$_{\rm vir}$ =1.8 (Seljak 2002). 
Our results are presented as squares for the 4:1 black hole growth scenario and 
pluses for 10:1 growth scenario only for those black holes that have completed 
the merger by redshift zero. The primary halo and its central SMBH fit the 
M$_{\rm DMH}$ vs. M$_{\rm BH}$ relation for 10:1 growth scenario while satellites 
and their central black hole broadly fit the extrapolated M$_{\rm DMH}$ vs. M$_{\rm BH}$
relation.}
\label{fig:roguehalo}
\end{center}
\end{figure}

\begin{figure}
\vspace{0.5in}
\begin{center}
\includegraphics [width=4.in,angle=0]{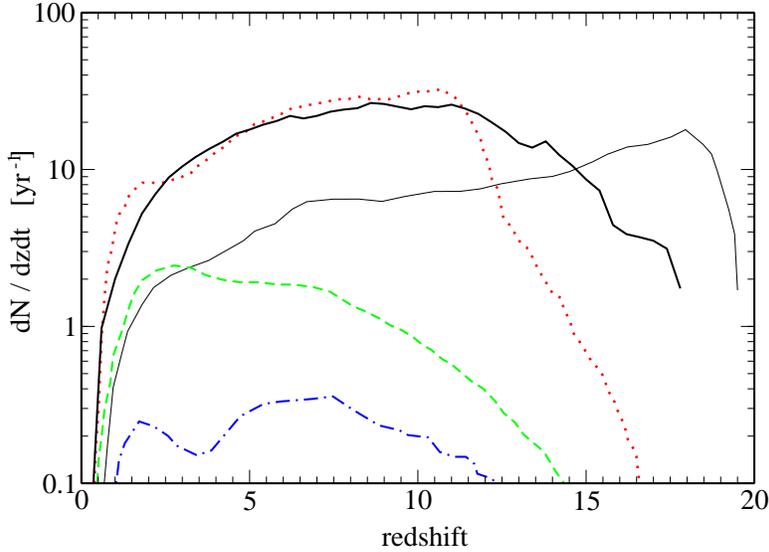}
\caption[Fig 16.]{Merger rates observed at z=0 as a function of redshift for four 
models described in Sesana et al. 2007, all with a Chandrasekhar prescription for
dynamical friction. The thin black line corresponds to massive DMHs 
(M$_{\rm DMH}$=10$^{11}$ - 10$^{15}$ $\Msun$) seeded with m$_{\rm BH}$$\sim$200$\Msun$ 
black holes at z=20; The dot-red line represents low mass halos (M$_{\rm DMH}$=10$^6$ - 10$^7$ $\Msun$) 
seeded with m$_{\rm BH}$$\sim$5$\times10^4\Msun$ at 15$\ltsim$z$\ltsim$20;
The blue-dashed line and black-dot line explore different redshift 
ranges for seeding black holes in halos: green: m$_{\rm BH}$=$10^4 - 10^5 \Msun$
at 15$\ltsim$z$\ltsim$20 and blue 18$\ltsim$z$\ltsim$20. Overplotted in the thick 
black is the black hole merger rate when dynamical friction is modeled with an n-body 
based estimate of the dynamical friction decay time (Boylan-Kolchin et al. 2008)}

\end{center}
\end{figure}


\begin{thebibliography}{dw}

\bibitem[Aguilar \& White(1986)]{1986ApJ...307...97A} Aguilar, L.~A., \& White, S.~D.~M.\ 1986, ApJ, 307, 97 
\bibitem[Alexander et al. 2005]{}Alexander, D.M., Smail, I., Bauer, F.E., \&
Chapman, S.C., Blain, A.W., Brandt, W.N., Ivison, R.J., 2005, Natur, 434, 738A
\bibitem[Aller $\&$ Richstone 2002]{}Aller M.C., Richstone D., 2002, AJ, 124, 3035
\bibitem[Abel et al. 2000]{Abel2000}Abel, T., Bryan, G., \&
Norman, M., 2000, ApJ, 540, 39     
\bibitem[Abel et al. 2002]{Abel2002}Abel, T., Bryan, G., \&
Norman, M., 2002, Sci, 295, 93A   
\bibitem[Adams et al. 2001]{}Adams, F.C., Graff, D.S., Richstone, D.O., 2001, ApJ, 551L, 31A
\bibitem[Baes et al. 2003]{}Baes, M., Buyle, P., Hau, G.K.T., Dejonghe, H., 2003, MNRAS, 341L, 44B
\bibitem[Barth et al. 2004]{}Barth, A.J., Ho, L.C., Rutledge, R.E., Sargent, W.L.W., \&
2004, ApJ, 607, 90B
\bibitem[Begelman et al. 1980]{1980Natur.287..307B} Begelman, M.C., \& 
Blandford, R.D., Rees, M.J., 1980, Natur, 287, 307 
\bibitem[Begelman \& Rees 1978]{}Begelman, M.C., Rees, M.J., 1978, MNRAS, 185, 847B
\bibitem[Bender et al. 2005]{}Bender, R., Kormendy, J., Bower, G., 2005, ApJ, 631, 280
\bibitem[Bennett et al. 2003]{2003ApJS..148....1B}Bennett, C.L., et al., 2003, ApJ, 148, 1 
\bibitem[Berczik et al. 2006]{}Berczik, P., Merritt, D., Spurzem, R., Bischof, H.P., \&
2006, ApJ, 642L, 21B   
\bibitem[Binney \& Tremaine 1987]{}Binney, J., Tremaine, S., 1987, Galactic Dynamics \&
(Princeton: Princeton Univ. Press)   
\bibitem[Bogdanovic et al. 2007]{}Bogdanovic, T., Reynolds, C.S., Miller, M.C.,  \& 
2007, ApJ, 661L, 147B 
\bibitem[Boylan-Kolchin et al. 2008]{}Boylan-Kolchin, M., Ma, C.P., Quataert, E., \&
MNRAS, 383, 93B   
\bibitem[Bromm et al. 1999]{1999ApJ...527L...5B}Bromm, V., Coppi, P.S.,   \&
Larson, R.B., 1999, ApJ, 527, L5 
\bibitem[Bromm \& Loeb 2003]{}Bromm, V., Loeb, A., 2003, ApJ, 596, 34B
\bibitem[Bromm \& Loeb 2004]{}Bromm, V., Loeb, A., 2004, NewA, 9, 353B    
\bibitem[Bullock et al. 2001]{}Bullock, J.S., Kollat, T.S., Sigad, Y., Somerville, R.S., \&
Kravtsov, A.V., Klypin, A.A., Primack, J.R., Dekel, A., 2001, MNRAS, 321, 559B   
\bibitem[Burkert \& Silk 2001]{}Burkert, A., Silk, J., 2001, ApJ, 554L, 151B
\bibitem[Campanelli et al. 2007]{}Campanelli, M., Lousto, C.O., Zlochower, Y., 
Merritt, D., 2007, gr.qc, 2133C   
\bibitem[Cattaneo et al. 1999]{}Cattaneo, A., Haehnelt, M.G., Rees, M.J., 1999, MNRAS, 308, 77C
\bibitem[Colpi et al. 1999]{}Colpi, M., Mayer, L., Governato, F., 1999, ApJ, 525, 720C   
\bibitem[Colpi et al.(2007)]{2007arXiv0710.5207C} Colpi, M., Dotti, M., 
Mayer, L., \& Kazantzidis, S.\ 2007, ArXiv e-prints, 710, arXiv:0710.5207 
\bibitem[Cowie et al. 2003]{}Cowie, L.L., Barger, A.J., Bautz, M.W., \& 
Brandt, W.N., Garmire, G.P., 2003, ApJ, 584, L5
\bibitem[David et al. 1987]{}David, L.P., Durisen, R.H., Cohn, H.N., 1987, ApJ, 313, 556D
\bibitem[Davis et al. 1985]{}Davis, M., Efstathiou, G., Frenk, C., White, S.D.M., \&
1985, ApJ, 292, 371
\bibitem[Dehnen et al. 2006]{Dehnen}Dehnen, W., McLaughlin, D.E. \& 
Sachania, J., 2006, MNRAS, 369, 1688
\bibitem[Di Matteo et al. 2003]{}Di Matteo, T., Croft, R.A.C., Springel, V., \& 
Hernquist, L., 2003, ApJ, 593, 56D
\bibitem[Di Matteo et al. 2005]{}Di Matteo, T., Springel, V., Hernquist, L., 2005, Natur, 433, 604D
\bibitem[Di Stefano 2007]{}Di Steffano, R., 2007, astro-ph/0712.3558
\bibitem[Dotti et al. 2007]{}Dotti, M., Colpi, M., Haardt, F., Lucio, M., 2007, MNRAS, 379, 956D   
\bibitem[Ebisuzaki et al. 1991]{}Ebisuzaki, T., Makino, J., Okumura, S.K., 1991, Natur, 354, 212E
\bibitem[Erickcek et al. 2006]{}Erickcek, A.L., Kamionkowski, M., \&
Benson, A.J., 2006, MNRAS, tmp, 940E
\bibitem[Escala et al. 2005]{Escala}Escala, A, Larson, R.B., Coppi, P.S.,  \&
Mardones, D., 2005, ApJ, 630, 152E
\bibitem[Fabbiano 1989]{Fabbiano}Fabbiano, G., 1989, ARA\&A, 27, 87
\bibitem[Fabbiano $\&$ White 2006]{}Fabbiano, G., White, N.E., 2006,  \&
in Compact Stellar X-Ray Sources, ed. W. H. G. Lewin \& M. van der Klis \&
(Cambridge: Cambridge Univ. Press), 475
\bibitem[Fan 2005]{}Fan, X., 2005, gbha.conf, 75F
\bibitem[Ferrarese \& Merritt 2000]{}Ferrarese, L., Merritt, D., 2000, ApJ, 539L, 9F
\bibitem[Ferrarese 2002]{}Ferrarese, L., 2002, ApJ, 578, 90F
\bibitem[Filippenko \& Ho 2003]{}Filippenko, A.V., Ho, L.C., 2003, ApJ, 588L, 13F
\bibitem[Gao]{}Gao, L., White, S.D.M., Jenkins, A., Frenk, C.S., Springel, V., \&
2005, MNRAS, 363, 379
\bibitem[Gebhardt et al. 2000]{}Gebhardt, K., et al. 2000, 539L, 13G
\bibitem[Gebhardt et al. 2005]{}Gebhardt, K., Rich, R.M., Ho, L,C., 2005, ApJ, 634, 1093G
\bibitem[Ghez et al. 2003]{}Gehz, A.M., Duchene, G., Matthews, K., 2003, ApJ, 586, 127
\bibitem[Ghez et al. 2005]{}Gehz, A.M., Salim, S., Hornstein, S.D., 2005, ApJ, 620, 744
\bibitem[Ghigna et al. 1998]{1998MNRAS.300..146G} Ghigna, S., Moore, B.,  \& 
Governato, F., Lake, G., Quinn, T., Stadel, J., 1998, MNRAS, 300, 146 
\bibitem[Gonzalez et al. 2007a]{}Gonzalez, J.A., Hannam, M., Sperhake, U., Brugmann, B.,
Husa, S., 2007, PhRvL, 98w1101G   
\bibitem[Gonzalez et al. 2007b]{}Gonzalez, J.A., Sperhake, U., Brugmann, B., Hannam, M., 
Husa, S., 2007, PhRvL, 98i1101G   
\bibitem[Granato et al. 2001]{}Granato, G.L., Silva, L., Monaco, P., Panuzzo, P., \& 
Salucci, P., De Zotti, G., Danese, L., 2001, MNRAS, 324, 757G
\bibitem[Green \& Ho 2004]{}Green, J.E., Ho, L.C., 2004, ApJ, 610, 722G
\bibitem[Greenstein \& Matthews 1963]{}Greenstein, J.L., Matthews, T.A, 1963, AJ, 68S, 279G
\bibitem[Haehnelt]{}Haehnelt, M.G., 1994, MNRAS, 269, 199
\bibitem[Haehnelt \& Kauffmann 2000]{}Haehnelt, M.G., Kauffmann, G., 2000, MNRAS, 318L, 35H
\bibitem[Haehnelt et al. 1998]{}Haehnelt, M.G., Natarajan, P., Rees, M.J., 1998, MNRAS, 300, 817H
\bibitem[Heckman et al. 2004]{}Heckman, T.M., Kauffmann, G., Brinchmann, J., \&
Charlot, S., Tremonti, C., White, S.D.M., 2004, ApJ, 613, 109
\bibitem[Heger \& Woosley 2002]{}Heger, A., Woosley, S.E., 2002, ApJ, 567, 532H   
\bibitem[Heger et al. 2003]{Heger}Heger et al., 2003, ApJ, 591, 288H     
\bibitem[Heggie et al. 2007]{}Heggie, D.C., Hut, P., Mineshige, S., Makino, J., \%
Baumgardt, H., 2007, PASJ, 59L, 11H   
\bibitem[Hernquist 1989]{1989Natur.340..687H}Hernquist, L., 1989, Natur, 340, 687 
\bibitem[Hernquist \& Mihos 1995]{1995ApJ...448...41H}Hernquist, L., \& Mihos, J.C., 1995, ApJ, 448, 41 
\bibitem[Herrmann et al. 2007]{}Herrmann, F., Hinder, I., Shoemaker, D., Laguna, P., \&
Matzner, R.A., 2007, ApJ, 661, 430H   
\bibitem[Ho 2008]{}Ho, L., 2008, astro-ph/0803.2268  
\bibitem[Ho et al. 1997]{1997ApJ...487..591H} Ho, L.C., Filippenko,  \& 
A.V., Sargent, W.L.W., 1997, ApJ, 487, 591 
\bibitem[Holley-Bockelmann et al.(2007)]{2007arXiv0707.1334H} 
Holley-Bockelmann, K., Gultekin, K., Shoemaker, D., 
\& Yunes, N.\ 2007, ArXiv e-prints, 707, arXiv:0707.1334 
\bibitem[Holley-Bockelmann \& Sigurdsson 2006]{}Holley-Bockelmann, K., \&
Sigurdsson, S., 2006, astro-ph, 1520H   
\bibitem[Holley-Bockelmann \& Richstone 1999]{}Holley-Bockelmann, K., \&
Richstone, D.O., 1999, ApJ, 517, 92H
\bibitem[Hopkins et al. 2005]{}Hopkins, P.F., Hernquist, L., Cox, T.J., Di Matteo, T., \& 
Martini, P., Robertson, B., Springel, V., 2005, ApJ, 630, 705H
\bibitem[Hu et al. 2006]{}Hu, J., Shen, Y., Lou, Y., Zhang, S., 2006, ApJ, 365, 345   
\bibitem[Islam et al. 2003]{Islam}Islam, R. R., Taylor, J. E., Silk, J., \&
2003, MNRAS, 340, 647I    
\bibitem[Islam et al. 2004]{Islam}Islam, R. R., Taylor, J. E., Silk, J., \&
2004, MNRAS, 354, 427I
\bibitem[Kauffmann \& Haehnelt 2000]{}Kauffmann, G., Haehnelt, M., 2000, MNRAS, 311, 576K
\bibitem[Kazantzidis et al. 2005]{}Kazantzidis, S., Mayer, L., Colpi, M., \&
Madau, P., Debattista, V.P., Wadsley, J., Stadel, J., Quinn, T., Moore, B., \&
2005, ApJ, 623L, 67K     
\bibitem[King et al. 2001]{King}King, A.R., Davies, M.B., Ward, M.J., Fabbiano, G. \& 
Elvis, M., 2001, ApJL, 552, 109
\bibitem[Koppitz et al. 2007]{}Koppitz, M., Pollney, D., Reisswig, C., Rezzolla, L., \&
Thornburg, J., Diener, P., Schnetter, E., 2007, gr.qc, 1163K   
\bibitem[Kormendy \& Richstone 1995]{}Kormendy, J., Richstone, D., 1995, ARA\&A, 33, 581K   
\bibitem[Koushiappas et al. 2004]{}Koushiappas, S.M., Bullock, J.S., Dekel, A., 2004, MNRAS, 354, 292K
\bibitem[Lacey \& Cole]{}Lacey, C., Cole, S., 1993, MNRAS, 262, 627
\bibitem[Lehmer et al. 2005]{}Lehmer, B.D., Brandt, W.N., Hornschemeier, A.E.,  \&
Alexander, D.M., Bauer, F.E., Koekemoer, A.M., Schneider, D.P., Steffen, A.T., \&
2006, AJ, 131, 2394L
\bibitem[Loeb \& Rasio 1994]{}Loeb, A., Rasio, F.A., 1994, ApJ, 432, 52L
\bibitem[Machacek et al. 2001]{}Machacek, M.E., Bryan, G.L., \&
Abel, T. 2001, ApJ, 548, 509
\bibitem[Mack et al.]{}Mack, K.J., Ostriker, J.P., Ricotti, M., astro-ph/0608642
\bibitem[Madau \& Rees 2001]{}Madau, P., Rees, M.J., 2001, ApJ, 551L, 27M
\bibitem[Makino 1997]{}Makino, J., 1997, ApJ, 478, 58M
\bibitem[Marconi et al. 2004]{2004MNRAS.351..169M} Marconi, A., Risaliti,   \& 
G., Gilli, R., Hunt, L.K., Maiolino, R., Salvati, M., 2004, MNRAS, 351, 169 
\bibitem[Mayer et al. 2007]{}Mayer, L., Kazantzidis, S., Madau, P., Colpi, M., \&
Quinn, T., Wadsley, J., 2007, Sci, 316, 1874M
\bibitem[Menou et al. 2001]{}Menou, K., Haiman, Z., Narayanan, V.K., \&
2001, ApJ, 558, 535
\bibitem[Merloni 2004]{}Merloni, A., 2004, MNRAS, 353, 1035
\bibitem[Merritt \& Ferrarese 2001]{}Merritt, D., Ferrarese, L., 2001, ApJ, 547, 140M
\bibitem[Merritt et al. 2004]{}Merritt, D., Milosavljevic, M., Favata, M., Hughes, S.A., 
Holz, D.E., 2004, ApJ, 607L, 9M  
\bibitem[Mihos \& Hernquist 1994]{}Mihos, J.C., Hernquist, L., 1994, ApJ, 425L, 13M
\bibitem[Mii \& Totani]{}Mii, H., Totani, T., 2005, ApJ, 628, 873
\bibitem[Miller \& Colbert 2004]{}Miller, M.C., Colbert, E.J.M., 2004, IJMPD, 13, 1M
\bibitem[Milosavljevic \$ Merritt 2001]{}Milosavljevic, M., Merritt, D., 2001, ApJ, 563, 34M
\bibitem[Milosavljevic \$ Merritt 2003]{}Milosavljevic, M., Merritt, D., 2003, ApJ, 596, 860M
\bibitem[Micic]{Micic}Micic, M., Abel, T., Sigurdsson, S., 2006, MNRAS, 372, 1540M 
\bibitem[Micic et al. 2007]{}Micic, M., Holley-Bockelmann, K., Sigurdsson, S., Abel, T., 
2007, MNRAS, 380, 1533M  
\bibitem[Monaco et al. 2000]{}Monaco, P., Salucci, P., Danese, L., 2000, MNRAS, 311, 279M
\bibitem[Nagashima et al. 2005]{}Nagashima, M., et al. 2005, ApJ, 634, 26N
\bibitem[Nakamura \& Umemura 2001]{}Nakamura, F., Umemura, M., 2001, ApJ, 548, 19N  
\bibitem[Navarro, Frenk \& White 1995]{}Navarro, J.F., Frenk, C.S., White, S.D.M., \&
1995, MNRAS, 275, 56N   
\bibitem[Omukai \& Yoshii 2003]{}Omukai, K., Yoshii, Y., 2003, ApJ, 599, 746O    
\bibitem[Omukai \& Palla 2003]{}Omukai, K., Palla, F., 2003, ApJ, 589, 677O    
\bibitem[Omukai \& Nishi 1998]{}Omukai, K., Nishi, R., 1998, ApJ, 508, 141O    
\bibitem[Portegies Zwart et al. 2004]{}Portegies Zwart, S.F., Baumgardt, H., Hut, P., \& 
Makino, J., McMillan, S.L.W., 2004, Natur, 428, 724P
\bibitem[Press \& Schechter]{PS}Press, W.H., Schechter P., 1974, ApJ, 187, 425
\bibitem[Ptak \& Colbert]{Ptak}Ptak, A., Colbert, E., 2004, ApJ, 606, 291
\bibitem[Quinlan 1996]{}Quinlan, G.D., 1996, NewA, 1, 35Q
\bibitem[Reed et al. 2007]{}Reed, D.S., Bower, R., Frenk, C.S., Jenkins, A, Theuns, T., \&
2007, MNRAS, 374, 2R
\bibitem[Rees 1984]{}Rees, M.J., 1984, ARA\&A, 22, 471R
\bibitem[Rhook \& Wyithe]{}Rhook, K.J., Wyithe, J.S.B., 2005, MNRAS, 361, 1145
\bibitem[Richstone(1976)]{1976ApJ...204..642R} Richstone, D.~O.\ 1976, 
ApJ, 204, 642 
\bibitem[Ripamonti et al. 2002]{}Ripamonti, E., Haardt, F., Ferrara, A., Colpi, M, \&
2002, MNRAS, 334, 401R    
\bibitem[Robertson et al. 2006]{}Robertson, B., Hernquist, L., Cox, T.J., Di Matteo, T., \& 
Hopkins, P.F., Martini, P., Springel, V., 2006, ApJ, 641, 90R
\bibitem[Sales et al. 2007]{2007MNRAS.379.1464S} Sales, L.V., Navarro,   \& 
J.F., Abadi, M.G., Steinmetz, M., 2007, MNRAS, 379, 1464 
\bibitem[Shankar et al. 2006]{}Shankar, F., Lapi, A., Salucci, P., De Zotti, G., Danese, L., \&
2006, ApJ, 643, 14S
\bibitem[Schneider et al. 2002]{Schneider}Schneider, R., Ferrara, A., \&
Natarajan, P., Omukai, K., 2002, ApJ, 571, 30S    
\bibitem[Schnittman \& Buonanno 2007]{}Schnittman, J.D., Buonanno, A., 2007, ApJ, 662L, 63S  
\bibitem[Seljak 2002]{}Seljak, U., 2002, MNRAS, 334, 797S
\bibitem[Sesana et al. 2004]{}Sesana, A., Haardt, F., Madau P., \&
Volonteri, M., 2004, ApJ, 611, 623
\bibitem[Sesana et al. 2007]{}Sesana, A., Volonteri, M., Haardt, F., \&
2007, astro-ph/0701556
\bibitem[Shakira \& Syunyaev 1973]{}Shakura, N.I., Syunyaev, R.A., 1973, A\&A, 24, 337S
\bibitem[Sigurdsson 2003]{}Sigurdsson, S., 2003, CQGra, 20S, 45S 
\bibitem[Soltan 1982]{}Soltan A., 1982, MNRAS, 200, 115S
\bibitem[Silk \& Rees 1998]{}Silk, J., Rees, M.J., 1998, A\&A, 331L, 1S
\bibitem[Spergel et al.2007]{2007ApJS..170..377S}Spergel, D.N., et al., 2007, ApJ, 170, 377 
\bibitem[Springel et al. 2001]{Springel}Springel, V., Yoshida, N., \&
White, S. D. M., 2001, NewA, 6, 79  
\bibitem[Springel 2000]{PGF}Springel, V., 2000, MPA   
\bibitem[Taffoni et al. 2003]{}Taffoni, G., Mayer, L., Colpi, M., Governato, F.,
2003, MNRAS, 341, 434T    
\bibitem[Tremaine et al. 2002]{}Tremaine, S., et al. 2002, ApJ, 574, 740T
\bibitem[van der Marel 2004]{}van der Marel, R.P., 2004, cbhg.symp, 37V
\bibitem[Volonteri et al. 2003]{Volonteri}Volonteri, M., Haardt, F., \&
Madau, P., 2003, ApJ, 582, 559   
\bibitem[Wang et al. 2006]{}Wang, J.M., Chen, Y.M., Zhang, F., 2006, ApJ, 647L, 17W
\bibitem[Weinberg 1989]{1989MNRAS.239..549W}Weinberg, M.D., 1989, MNRAS, 239, 549 
\bibitem[Wise  \&  Abel 2005]{Wise}Wise, J., H., Abel, T., 2005, ApJ, 629, 615W   
\bibitem[Wyithe \& Loeb 2003]{}Wyithe, J.S.B., Loeb, A., 2003, ApJ, 590, 691
\bibitem[Wyithe \& Loeb 2005]{}Wyithe, J.S.B., Loeb, A., 2005, ApJ, 634, 910W


\end{thebibliography}
\end{document}